\documentclass[fleqn,usenatbib]{mnras}

\usepackage{newtxtext,newtxmath}
\usepackage[T1]{fontenc}

\DeclareRobustCommand{\VAN}[3]{#2}
\let\VANthebibliography\thebibliography
\def\thebibliography{\DeclareRobustCommand{\VAN}[3]{##3}\VANthebibliography}

\usepackage{graphicx}	% Including figure files
\usepackage{amsmath}	% Advanced maths commands
\usepackage{pdflscape}	% Landscape pages
\setcitestyle{notesep={; }}
\title[FLAMINGO galaxy clusters]{The FLAMINGO Project: Galaxy clusters in comparison to X-ray observations}

\author[Joey Braspenning et al.]{
Joey Braspenning,$^{1}$\thanks{E-mail: braspenning@strw.leidenuniv.nl}
Joop Schaye,$^{1}$
Matthieu Schaller,$^{2, 1}$
Ian G. McCarthy,$^{3}$
Scott T. Kay,$^{4}$
\newauthor
John C. Helly,$^{5}$
Roi Kugel,$^{1}$
Willem Elbers,$^{5}$
Carlos S. Frenk,$^{5}$
Juliana Kwan,$^{6}$
Jaime Salcido,$^{3}$
\newauthor
Marcel P. van Daalen,$^{1}$
Bert Vandenbroucke$^{1}$
\\
$^{1}$Leiden Observatory, Leiden University, PO Box 9513, 2300 RA Leiden, the Netherlands\\
$^{2}$Lorentz Institute for Theoretical Physics, Leiden University, PO box 9506, 2300 RA Leiden, the Netherlands\\
$^{3}$Astrophysics Research Institute, Liverpool John Moores University, Liverpool L3 5RF, UK\\
$^{4}$Jodrell Bank Centre for Astrophysics, Department of Physics and Astronomy, The University of Manchester, Oxford Road, Manchester M13 9PL, UK\\
$^{5}$Institute for Computational Cosmology, Department of Physics, University of Durham, South Road, Durham, DH1 3LE, UK\\
$^{6}$Department of Applied Mathematics and Theoretical Physics, University of Cambridge, Wilberforce Road, Cambridge, CB3 0WA, UK
}

\date{Accepted XXX. Received YYY; in original form ZZZ}

\pubyear{2024}

\begin{document}
\label{firstpage}
\pagerange{\pageref{firstpage}--\pageref{lastpage}}
\maketitle

% Abstract of the paper
\begin{abstract}
Galaxy clusters are important probes for both cosmology and galaxy formation physics. We test the cosmological, hydrodynamical FLAMINGO simulations by comparing to observations of the gaseous properties of clusters measured from X-ray observations. FLAMINGO contains unprecedented numbers of massive galaxy groups ($>10^6$) and clusters ($>10^5$) and includes variations in both cosmology and galaxy formation physics. We predict the evolution of cluster scaling relations as well as radial profiles of the temperature, density, pressure, entropy, and metallicity for different masses and redshifts. We show that the differences between volume-, and X-ray-weighting of particles in the simulations, and between cool-core non cool-core samples, are similar in size as the differences between simulations for which the stellar and AGN feedback has been calibrated to produce significantly different gas fractions. Compared to thermally-driven AGN feedback, kinetic jet feedback calibrated to produce the same gas fraction at $R_{\rm 500c}$ yields a hotter core with higher entropies and lower densities, which translates into a smaller fraction of cool-core clusters. Stronger feedback, calibrated to produce lower gas fractions and hence lower gas densities, results in higher temperatures, entropies, and metallicities, but lower pressures. The scaling relations and thermodynamic profiles show almost no evolution with respect to self-similar expectations, except for the metallicity decreasing with redshift. We find that the temperature, density, pressure, and entropy profiles of clusters in the fiducial FLAMINGO simulation are in excellent agreement with observations, while the metallicities in the core are too high.

\end{abstract}

% Select between one and six entries from the list of approved keywords.
% Don't make up new ones.

%%%%%%%%%%%%%%%%%%%%%%%%%%%%%%%%%%%%%%%%%%%%%%%%%%

%%%%%%%%%%%%%%%%% BODY OF PAPER %%%%%%%%%%%%%%%%%%

\section{Introduction}

The largest collapsed structures in our Universe, galaxy clusters, are an excellent probe of some of the most violent events we can observe, such as mergers with other clusters and feedback from accretion onto the most massive black holes. Furthermore, their abundance and clustering are important probes for cosmology. Because of their large mass, they are, unfortunately, quite rare, and statistical samples are much smaller than for lower-mass objects like galaxies.

Galaxy clusters are typically found in optical surveys \citep[e.g][]{Takey2013, Rykoff2014}, X-ray surveys \citep[e.g.][]{Bohringer2004, Ghirardini2019, Lovisari2017, Liu2022}, or Sunyaev-Zeldovich observations \citep[e.g.][]{Planck2016, Bleem2015, Hilton2021}. Optical surveys look for overdense areas of the sky, where many galaxies are clustered around a central bigger object, the brightest cluster galaxy (BCG). The number of such galaxies, called satellites, is a good measure of the mass of the cluster \citep[e.g.][]{Pereira2018}. 

Because clusters contain large amounts of hot gas, X-ray observations provide an alternative way to identify clusters. X-ray observations probe the intracluster medium (ICM), unlike longer wavelengths (such as the optical) which only identify galaxies. By imaging the ICM, X-ray observations can be used to construct surface brightness profiles \citep[e.g.][]{Neumann1999}, and the (deprojected) temperature, metallicity, and density profiles can be derived by fitting the spectrum \citep[e.g.][]{Vikhlinin2006, Sun2009}. These can then be combined to measure entropy and pressure profiles \citep[e.g.][]{Ponman2003, Arnaud2010}. 

The density and temperature profiles can be used to derive the hydrostatic mass of clusters through the equation of hydrostatic equilibrium \citep[e.g.][]{Ettori2013}. However, such masses tend to be biased, as clusters are never truly in hydrostatic equilibrium, and there may be significant non-thermal pressure. This ratio between the measured hydrostatic mass and the true mass, is called the hydrostatic bias. Comparing predicted hydrostatic masses with true masses in simulations leads to a hydrostatic bias of $\approx 0.8-0.9$ \citep[e.g.][]{Lebrun2014, Biffi2016, Barnes2021, Gianfagna2021, Jennings2023}. Alternatively, observational comparisons between X-ray inferred masses and masses obtained from weak lensing can be made \citep[e.g.][]{Mahdavi2013, vonderlinden2014, Henson2017}, giving a slightly larger hydrostatic bias ($\approx 0.73$) \citep[e.g.][]{Munoz2023, Hoekstra2015}.

Even though clusters deviate from perfect hydrostatic equilibrium, they are still the largest collections of particles in the Universe showing behaviour which can, at least in part, be described by such simple physics. As a result, it is of particular interest to compare scaling relations of different physical quantities associated with clusters. Common scaling relations in the literature are the mass--X-ray-luminosity relation \citep[e.g.][]{Lovisari2015, Lovisari2020, Gaspari2019}, mass--temperature relation \citep[e.g.][]{Pearson2017}, and the temperature--X-ray-luminosity relation \citep[e.g.][]{Pratt2009, Migkas2020}. 

At the lower mass end, lower temperatures lead to more metal emission lines, boosting the integrated X-ray luminosity over the thermal bremsstrahlung expectation \citep[e.g.][]{Lovisari2021}. However, due to non-gravitational processes such as AGN feedback pushing out gas from the centers of groups, the scaling relations do not show any such boost towards the low mass and low temperature end. The gas fraction within $R_{\rm 2500c}$\footnote{$r_{\Delta\rm c}$ is the radius within which the mean internal density is $\Delta$ times the critical density of the universe. The mass contained within $r_{\Delta\rm c}$ is denoted $M_{\Delta\rm c}$.} is observed to be strongly mass dependent, whereas it is almost mass independent in the range $R_{\rm 2500c}-R_{\rm 500c}$, suggesting that gas is pushed out of the centers of groups, but does not escape the system \citep{Sun2009}. The mass dependence of the mass--X-ray-luminosity relation also depends on the X-ray band within which the flux is integrated, as some lines may or may not be captured. In particular, it has been shown that the $0.5-2.0~\mathrm{keV}$ band is fairly insensitive to the metallicity, but the wider $0.1-2.4~\mathrm{keV}$ band, as well as the bolometric luminosity, are sensitive to the metallicity and the assumed element abundances for lower temperatures ($\mathrm{< 1 ~ keV}$) \citep[e.g.][]{Lovisari2021}.

Whereas scaling relations characterize clusters with single quantities such as mass, luminosity, or temperature, radial profiles give insight into the physical processes at play at different distances from the cluster centre. They are of particular interest when comparing simulations with observations, as agreement with observations would be a strong indication that the physical processes implemented in those simulations yield, on the scales represented by galaxy clusters, conditions that are realistic. Direct comparisons are, however, difficult. For example, observed density, temperature, and metallicity profiles are derived from fitting models to the X-ray spectrum of clusters within radial bins, whereas in simulations mass- or volume-weighted quantities are often used \citep[e.g.][]{Lehle2023, Li2023, Nelson2023, Pakmor2023, Towler2023}, but there are already long standing efforts to create mock observations and measure cluster properties directly from those \citep[e.g.][]{Nagai2007, Rasia2012, Mccarthy2017, Robson2023}. Because the X-ray luminosity, and hence the contribution to the total observed spectrum of a region, scales with the square of the gas density, denser regions contribute more to the quantities inferred from observations than they would in mass- or volume-weighted simulations. For the inferred temperature there is an empirical formulation, the spectroscopic-like temperature, to mimic the temperature derived from observations in the massive (\textgreater 3keV) cluster regime \citep{Mazzotta2004}.

Observational X-ray selection tends to be biased towards high surface brightness clusters, especially at higher redshifts \citep[e.g.][]{Eckert2011}. High surface brightness clusters are typically core-dominated, and classified as cool-core clusters since their temperature profile shows a central drop. The bias decreases but is still present for Sunyaev-Zel{\textquoteright}dovich (SZ) selected samples \citep[e.g.][]{Lin2015, Rossetti2017}. Furthermore, observational analyses have to deproject the X-ray maps to obtain 3D profiles, which adds significant uncertainty as it relies on the assumed sphericity of the cluster and limited thermal structure \citep[e.g.][]{Bartalucci2023}. In particular, this approach tends to overestimate central temperatures \citep{Lakhchaura2016}.

Due to their rarity, even in simulations the number of galaxy clusters tends to be low, as enormous volumes are required to obtain a large population. Here we analyze the FLAMINGO cosmological hydrodynamical simulations \citep{schaye2023, Kugel2023}, which include the largest full physics simulation run to $z=0$ to date. FLAMINGO includes different numerical resolutions and volumes up to $\mathrm{2.8~Gpc}$ on a side. The subgrid feedback in the fiducial model has been calibrated to reproduce the observed low-redshift galaxy stellar mass function and cluster gas fractions, while model variations produce different mass functions and/or gas fractions, use jet-like kinetic AGN feedback instead of thermally-driven AGN feedback, or assume different cosmologies. The unprecedented combination of volume and resolution offers a very large number of resolved clusters to compare with observational samples, e.g.\ more than two million haloes with mass $M_{500\rm c} > \mathrm{10^{13}~M_{\odot}}$ at $z=0$. Feedback and cosmology variations allow the study of the relative importance of different effects, and elucidate the physics driving the evolution and observational appearance of these rare objects.

In this paper, we analyse the scaling relations and thermodynamic profiles of clusters in the FLAMINGO simulations and compare them with observations. All runs have the same resolution, with the exception of convergence tests in the Appendix, for model comparisons we use the ($1~\mathrm{Gpc}$)$^3$ volumes, all other results are based off the ($2.8~\mathrm{Gpc}$)$^3$ volume. We will show how the results from the simulations depend on sample composition in mass, redshift, and cool-core fraction, as well as algorithmic choices, and how these choices affect the match with observational results. First, we introduce the FLAMINGO simulations, halo selection, and X-ray calculation in Section \ref{methods}. We then focus on the effect of model variations in FLAMINGO on the scaling relations in Section \ref{sec:scaling_relations_models}. We study thermodynamic profiles, and their susceptibility to different weighting schemes and cluster selections in Section \ref{sec:profiles}. Finally, we discuss our results and offer conclusions in Section \ref{sec:conclusion}.

\section{Methods} \label{methods}
\subsection{Simulations overview}
FLAMINGO (Full-hydro large-scale structure simulations with all-sky mapping for the interpretation of next generation observations) is a large suite of hydrodynamical cosmological simulations, covering enormous cosmic volumes. The flagship run comprises a region of $\mathrm{(2.8~Gpc)^3}$, which is ideal for the statistical studies of clusters. The simulations are described in detail in \citet{schaye2023}. A unique feature is the machine learning-aided calibration of the stellar and AGN feedback to reproduce the low-redshift cluster gas fractions at $R_{\rm 500c}$ and the $z=0$ galaxy stellar mass function \citep{Kugel2023}. 

Variations on the fiducial physical model in $\mathrm{(1~Gpc)^3}$ volumes are made by shifting the observed gas fractions (or galaxy stellar mass function) up or down with a multiple of their uncertainty ($\sigma$) (see Table~\ref{table:variations}) and recalibrating the model to fit those new data points.
For reference, at $M_\textrm{500c}=10^{14}~{\rm M}_\odot$ models fgas$+2\sigma$ and fgas$-8\sigma$ have gas mass fractions of $\approx 0.10$ and 0.05, respectively, while the fiducial model has 0.08 in agreement with observations. We note that if we interpret the gas fraction variations as horizontal shifts in the plot of the gas fraction as a function of mass, then the mass $M_\textrm{500c}$ where the gas fraction equals half the cosmic baryon fraction ($f_{\rm gas} \approx 0.08$), is about 0.2~dex lower and 0.5~dex higher for the models with the highest (fgas$+2\sigma$) and lowest (fgas$-8\sigma$) gas fractions, respectively. Besides the feedback variations, FLAMINGO also includes cosmology variations in $\mathrm{1~Gpc}$ volumes. We find that the cosmology variations have a negligible impact on the cluster properties investigated here (see \citet{schaye2023} for a comparison of cluster scaling relations), and hence we do not include them in this paper. 

In this work we make use of the fiducial resolution FLAMINGO simulations ($m_{\rm gas} = 1.07 \times 10^9 ~ \mathrm{M_\odot}$), and use the high ($m_{\rm gas} = 1.34 \times 10^8 ~ \mathrm{M_\odot}$), and low ($m_{\rm gas} = 8.56 \times 10^9 ~ \mathrm{M_\odot}$) resolution simulations for convergence testing in Appendix \ref{sec:appendix_convergence}, where we show the convergence is generally excellent for the thermodynamic profiles. The largest volume and all the model variations use the fiducial resolution. An overview of the simulations used in this work is presented in Table \ref{table:variations}.

The mass ranges over which the simulations have been calibrated depend on the resolution and the box size used for the calibration runs. The m9 simulations have been calibrated to the $z=0$ galaxy stellar mass function in the stellar mass range $10^{9.92}-10^{11.5}~\text{M}_\odot$ and low-$z$ cluster gas fractions in the mass range $M_\text{500c} = 10^{13.5} - 10^{14.36} ~\text{M}_\odot$. The lower (m10) and higher (m8) resolution simulations have the same upper limit for the stellar mass range but lower limit of, respectively, $10^{11.17}~\text{M}_\odot$ and $10^{8.67}~\text{M}_\odot$. The cluster gas fraction are calibrated from the same lower limit, but have different upper limits, respectively, $10^{14.53} ~\text{M}_\odot$ and $10^{13.73} ~\text{M}_\odot$ \citep{Kugel2023}.

FLAMINGO uses the open source simulation code \textsc{swift} \citep{Schaller2023} and the \textsc{sphenix} Smooth Particle Hydrodynamics (SPH) scheme \citep{Borrow2022}. Neutrinos are included as particles (with a summed mass of $\mathrm{0.06~eV}$ using the $\delta$f method \citep{Elbers2021}). Initial conditions are generated with a modified version of \textsc{monofonic} \citep{Hahn2021, Elbers2022}, and the default cosmology is the `3x2pt + all external constraints' from the dark energy survey year 3 ($\Omega_{\rm m} = 0.306$, $\Omega_{\rm b} = 0.0486$, $\sigma_8 = 0.807$, $\mathrm{H_0} = 68.1$, $n_{\rm s} = 0.967$; \citealt{Abbott2022}).

The FLAMINGO model includes subgrid implementations of radiative cooling \citep{ploeckinger2020}, star formation \citep{Schaye2008}, stellar mass loss\footnote{Stellar yields tabulated by \citet{Marigo2001, Portinari1998} and \citet{Thielemann2003} are used. The massive star yields of C, Mg and Fe are multiplied by factors of 0.5, 2 and 0.5, respectively, as in \citet{Wiersma_stellarmassloss2009}.} \citep{Wiersma_stellarmassloss2009, Schaye2015}, supernova feedback \citep{Schaye2008, Chaikin2022}, seeding and growth of black holes \citep{Springel2005, Bahe2022}, and thermally-driven AGN feedback \citep{BoothSchaye2009}. Additionally, two simulations variations use kinetic jet feedback from AGN \citep{Husko2022}.

\begin{figure*}
    \centering
    \includegraphics[width = \linewidth]{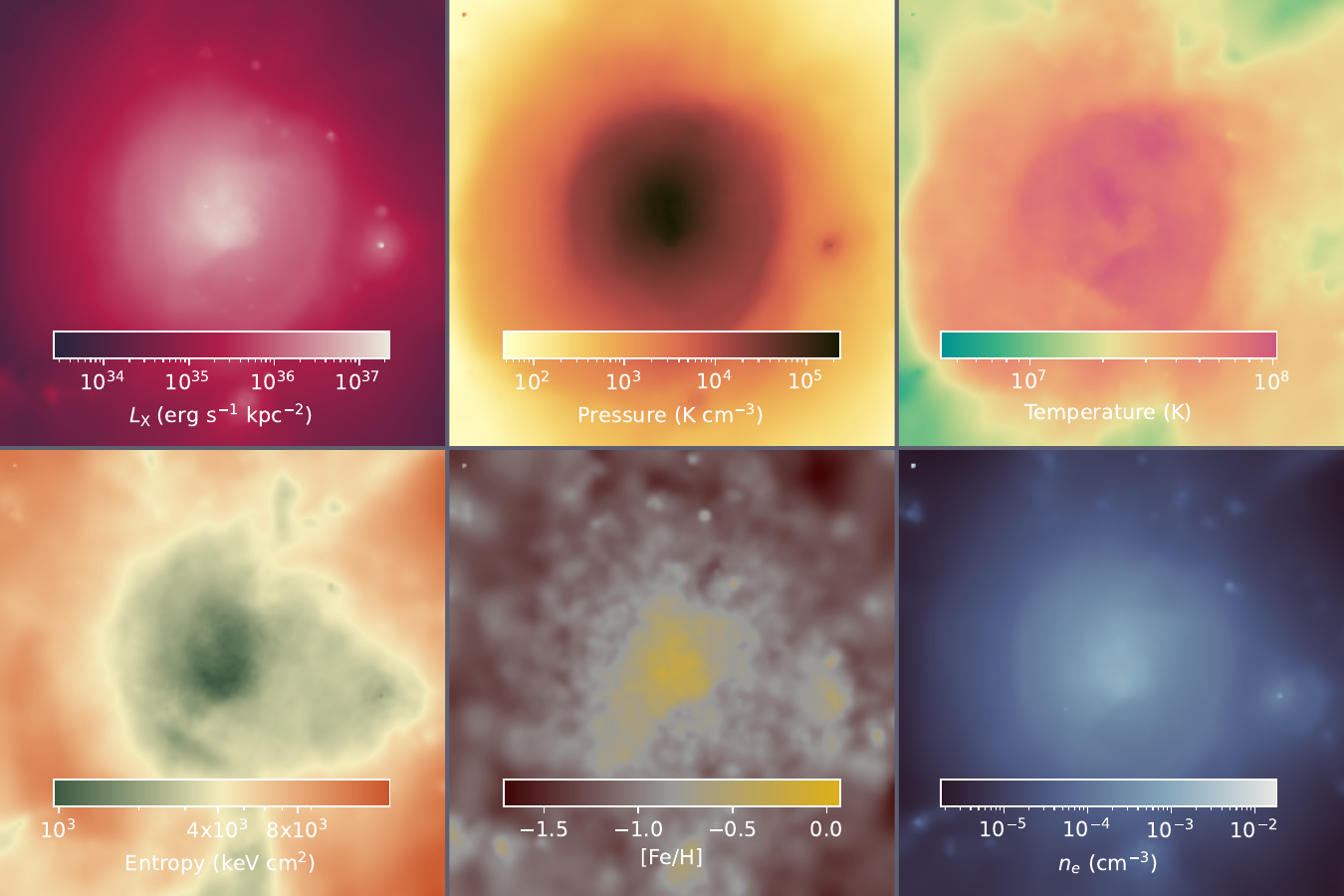}
    \caption{Projected maps for a massive cluster, $M_{\rm 500c} = 10^{15.03}~ \mathrm{M}_{\odot}$, of 4$R_{\rm 500c} (6.3~\mathrm{Mpc}$) on a side from the L1\textunderscore m9 simulation at $z=0$. Clockwise from the top left the different panels show projected X-ray surface luminosity density in the ROSAT band ($\mathrm{0.5-2.0 ~keV}$), pressure, temperature, free electron density, iron abundance and gas entropy, where the last five panels show mass weighted means.}
    \label{fig:graphical_cluster}
\end{figure*}

An example of a FLAMINGO halo is shown in Fig.~\ref{fig:graphical_cluster}, where we show mass-weighted projected thermodynamic quantities.

\subsection{X-ray luminosities}
We generate tables of collisional- and photo-ionized X-ray model spectra using Cloudy (version 17.02; \citealt{Ferland2017}). Contrary to the common practice of using separate packages for the radiative cooling rates used in the simulation and for X-ray spectra generated in post-processing, we self-consistently use Cloudy for both. To compute the X-ray spectra we follow the methods used by \citet[hereafter PS20]{ploeckinger2020} for the radiative cooling rates used in FLAMINGO. We use the same density ($10^{-8}~\mathrm{cm^{-3}} \leq n_{\rm H} \leq 10^{6} ~\mathrm{cm^{-3}}$), redshift ($0 \leq z \leq 9$) and metallicity (primordial to $3Z_{\odot}$) ranges, but add information about the X-ray luminosity. We limit the analysis to gas with temperatures $10^5 ~\mathrm{K} \leq T \leq 10^{9.5} ~\mathrm{K}$, as this is the relevant regime for X-ray clusters. In line with PS20, we use the UV and X-ray background from \citet{FG20} modified at $z > 3$ to match the effective photo-ionisation and photo-heating rates before helium reionisation.

To compute the element-by-element emissivities, we generate differential spectra isolating the contribution of a single metal (analogous to the \citealt{wiersma2009} approach). We do this for 9 metals (C, N, O, Ne, Mg, Si, Ca, S, and Fe), 7 of which are explicitly traced by FLAMINGO (Ca and S are assumed to trace Si with mass ratios of 0.094 and 0.605)\footnote{This is consistent with our radiative cooling prescription, see \citet{schaye2023}.}. We make use of the solar abundance table of \citet{Asplund2009}. A final spectrum can then be generated by taking the base spectrum, which lacks all 9 metals, and putting their respective contributions back in according to their individual abundances. This allows for non-solar relative abundances to be captured. These differential spectra are generated for every grid-point in metallicity, density, temperature, and redshift. Helium is treated differently because its contribution to the free electron abundance is non-negligible, every metallicity bin corresponds to a different helium abundance, and for every such bin the differential spectra are generated.

Metals which are not taken out individually are always assumed to be at the metallicity corresponding to the helium fraction assuming a solar abundance pattern\footnote{FLAMINGO computed X-ray luminosities with all non-traced metals always set to solar metallicity, for future work this has been resolved. However, it makes a negligible difference for groups and clusters.}.

We employ a single zone cloudy model, which is appropriate for unshielded or ionised gas, and analogous to the \citet{wiersma2009} approach. PS20 account for self-shielding in dense, cool gas, but self-shielding is not relevant for X-ray clusters. A cosmic ray background is used with the default value from Cloudy. PS20 change the value of the cosmic ray background depending on the shielding column, but this is again not relevant for X-ray clusters (increasing the cosmic ray background by an order of magnitude has a $\mathrm{< 0.3\%}$ effect on the result and only at the lowest temperatures). No grains are used, since we are only working in the $\mathrm{T > 10^5\, K}$ regime. This is a safe assumption as figure~7 of PS20 shows that there is assumed to be no dust at these temperatures.

\begin{table*} 
	\centering
	\caption{Hydrodynamical simulation runs. The columns list the simulation name, the number of standard deviations by which the observed stellar masses are shifted before calibration $\Delta m_\ast$, the number of standard deviations by which the observed cluster gas fractions are shifted before calibration $\Delta f_\text{gas}$, the comoving box side length $L$, the AGN feedback implementation (thermal or jets), the number of baryonic particles $N_\text{b}$ (which equals the number of cold dark matter particles $N_\text{CDM})$, the number of neutrino particles $N_\nu$, the initial mean baryonic particle mass $m_\text{b}$, the mean cold dark matter particle mass $m_\text{CDM}$, the comoving gravitational softening length $\epsilon_\text{com}$, and the maximum proper gravitational softening length $\epsilon_\text{prop}$.}
    \label{table:variations}
	\begin{tabular}{lrrlccccccc} 
		\hline
		Identifier & $\Delta m_\ast$ & $\Delta f_\text{gas}$ & AGN & $L$ & $N_\text{b}$ & $N_\nu$ & $m_\text{g}$ & $m_\text{CDM}$ & $\epsilon_\text{com}$ & $\epsilon_\text{prop}$ \\
		           & ($\sigma$) & ($\sigma$) && (cGpc) &&& ($\mathrm{M_\odot}$) & ($\mathrm{M_\odot}$)  & (ckpc) & (pkpc) \\
		\hline
		L2p8\_m9             & 0 & 0 & thermal &2.8& $5040^3$ & $2800^3$ & $1.07\times 10^9$ & $5.65\times 10^9$    & 22.3  & 5.70\\
		L1\_m8             & 0 & 0 & thermal & 1 & $3600^3$ & $2000^3$ & $1.34\times 10^8$ & $7.06\times 10^8$    & 11.2  & 2.85\\
		L1\_m9             & 0 & 0 & thermal & 1 & $1800^3$ & $1000^3$ & $1.07\times 10^9$ & $5.65\times 10^9$    & 22.3  & 5.70\\
		L1\_m10              & 0 & 0 & thermal & 1 &  $900^3$  & $500^3$ & $8.56\times 10^9$ & $4.52\times 10^{10}$ & 44.6 & 11.40\\
		fgas$+2\sigma$\_L1\_m9     & 0 & $+2$ & thermal & 1 & $1800^3$ & $1000^3$ & $1.07\times 10^9$ & $5.65\times 10^9$ & 22.3  & 5.70\\
		fgas$-2\sigma$\_L1\_m9     & 0 & $-2$ & thermal & 1 & $1800^3$ & $1000^3$ & $1.07\times 10^9$ & $5.65\times 10^9$    & 22.3  & 5.70\\
		fgas$-4\sigma$\_L1\_m9     & 0 & $-4$ & thermal & 1 & $1800^3$ & $1000^3$ & $1.07\times 10^9$ & $5.65\times 10^9$    & 22.3  & 5.70\\
		fgas$-8\sigma$\_L1\_m9     & 0 & $-8$ & thermal & 1 & $1800^3$ & $1000^3$ & $1.07\times 10^9$ & $5.65\times 10^9$    & 22.3  & 5.70\\
		M*$-\sigma$\_L1\_m9     & $-1$ & 0 & thermal & 1 & $1800^3$ & $1000^3$ & $1.07\times 10^9$ & $5.65\times 10^9$    & 22.3  & 5.70\\
		M*$-\sigma$\_fgas$-4\sigma$\_L1\_m9     & $-1$ & $-4$ & thermal & 1 & $1800^3$ & $1000^3$ & $1.07\times 10^9$ & $5.65\times 10^9$    & 22.3  & 5.70\\
		Jet\_L1\_m9             & 0 & 0 & jets & 1 & $1800^3$ & $1000^3$ & $1.07\times 10^9$ & $5.65\times 10^9$    & 22.3  & 5.70\\
		Jet\_fgas$-4\sigma$\_L1\_m9             & 0 & $-4$ & jets & 1 & $1800^3$ & $1000^3$ & $1.07\times 10^9$ & $5.65\times 10^9$    & 22.3  & 5.70\\
		\hline
	\end{tabular}
\end{table*}

\subsection{X-ray table generation}
At each redshift, the spectra computed by Cloudy are integrated over the $z=0$ observer band. The table stores both the photon emissivity $\mathrm{[s^{-1} ~cm^{3}]}$, and energy emissivity $\mathrm{[erg~s^{-1}~cm^{3}]}$. For the energy version of the table the last column of the Cloudy output file \verb+"diffuse.dat"+ is used. This contains the total diffuse emisssivity $4\pi \nu j_\nu$ ($\mathrm{erg\, cm^{-3}\, s^{-1}}$) (continuum + lines), where $\nu$ is the frequency and $j_\nu$ the specific intensity. For the photon count version of the table the output from the Cloudy files \verb+"XSPEC_diffuse_reflected.FITS"+, \verb+"XSPEC_diffuse.FITS"+, \verb+"XSPEC_lines.FITS"+ and \verb+"XSPEC_lines_reflected.FITS"+ (all in units of $\mathrm{photons\, s^{-1}\, cm^{-2}\, bin^{-1}}$) is summed. To obtain a photon emissivity from this flux, the value is divided by the depth of the shell.

Calling the emissivity spectrum obtained from Cloudy $\varepsilon_{\nu}$ and the midpoint energy of the spectral bins $E_{\nu}$, the following integral is performed.
\begin{equation}
    J_{[E_{-} - E_{+}]} = \int_{E_{-}}^{E_{+}} \frac{\varepsilon_\nu}{E_\nu} \mathrm{d} E_\nu ,
\end{equation}
where $J_{[E_{-} - E_{+}]}$ ($\mathrm{erg \, cm^{-3} s^{-1}}$) is the total emissivity over the energy range $E_{-}$ to $E_{+}$. The boundary values of this energy range are fixed for an observer at $z=0$, and hence they are defined as
\begin{align}
    E_{-} &= E_{-, z=0}\; (1+z) \, ,\\
    E_{+} &= E_{+, z=0}\; (1+z) \, .
\end{align}
We create integrated broadbands covering energy ranges corresponding to observational campaigns: 
\begin{itemize}
    \item \textbf{ROSAT} 0.5 - 2.0 keV
    \item \textbf{eROSITA-low} 0.2 - 2.3 keV
    \item \textbf{eROSITA-high} 2.3 - 8.0 keV
\end{itemize}
This work makes use only of the ROSAT band. For each grid point in the tables, the emissivity of these bands is integrated using the trapezoidal rule. Finally, to account for the large dynamic range in emissivities due to the scaling with the square of the density, we divide all the emissivities by the square of the hydrogen density
\begin{equation}
    \hat{J}_{[E_{-} - E_{+}]} = J_{[E_{-} - E_{+}]} / n_{\rm H}^2 \,
\end{equation}
giving units of $\mathrm{[erg~s^{-1}~cm^{3}]}$.
We store $\log_{10} \left(\hat{J}_{[E_{-} - E_{+}]} \right)$ in the tables.

\subsection{X-ray interpolation}
With the Cloudy-generated table we interpolate the logarithmic X-ray broadband linearly to the logarithms of the density, temperature and elemental abundances of the simulation resolution elements (particles in the case of FLAMINGO), following the five steps below.
\paragraph*{Step 1} The density bin (\verb+idx_n+) and temperature bin (\verb+idx_T+) corresponding to the resolution element are found and the absolute offset from the bin centres (\verb+dx_T+, \verb+dx_n+).
\paragraph*{Step 2} The abundances and ratio-to-solar are computed. The abundance of each element (helium, carbon, nitrogen, oxygen, neon, magnesium, silicon and iron) is computed using
\begin{equation}
    \text{A}_i = \frac{X_i}{X_{\rm H}} \frac{1}{m_i / m_{\rm H}},
\end{equation}
where $X_i$ and $X_{\rm H}$ are the element and hydrogen mass fractions taken from the simulation and $m_i$ the element atomic mass in units of the proton mass.

Ratio-to-solar ($R$) is the ratio of the element abundance from the simulation to the solar abundance
\begin{equation}
    R_i = \frac{\text{A}_i}{\text{A}_{i, \odot}}.
\end{equation}
Note that the simulations predict absolute abundances and that the results therefore do not depend on the assumed solar abundances of the elements tracked by the simulations provided the same solar abundances are applied throughout the analysis. 

\paragraph*{Step 3} The appropriate helium bin and the relative offset to that bin centre are found, using the abundance from Step 2.

\paragraph*{Step 4} The interpolation over the helium, temperature, density and redshift axes of the table is performed for all 9 differential values of the luminosity with a single metal missing (9 entries with only a single metal present, 1 entry with no metals present)
\paragraph*{Step 5}
The contributions from all metals are added based on their ratio-to-solar. With $J_{\text{no metals}}$ representing the interpolated emissivity with \emph{all} 9 tracked metals removed, and $J_i$ the interpolated emissivity where metal $i$ is present, this is computed as
\begin{equation}
    \log_{\rm 10} J_{\text{final}} = \log_{\rm 10} J_{\text{no metals}} + \sum_i \log_{\rm 10} J_i  \times R_i
\end{equation}

\subsection{Halo selection} \label{sec:halo_selection}
In this work we select all haloes for which the mass,
$M_{\rm 500c}$, is larger than $10^{13} ~\mathrm{M_\odot}$ from the halo catalogs of the FLAMINGO simulations. These are generated using the 6D Friends-Of-Friends subhalo finder Velociraptor \citep{Elahi2019vr}. Physical quantities are then computed using the Spherical Overdensity and Aperture Processor (SOAP), a tool developed for the FLAMINGO project. SOAP computes (sub)halo properties using as an input only the (sub)halo centers and particle membership from the halo finder.

For the fiducial model, the L2p8\_m9 and L1\_m9 simulations have, respectively, 2,031,904 and 92,015 haloes with $M_{\text{500c}} > 10^{13}~ \mathrm{M_\odot}$, 113,168 and 5,164 haloes with $M_{\text{500c}} > 10^{14}~ M_\odot$, and 461 and 15 haloes with $M_{\text{500c}} > 10^{15}~ \mathrm{M_\odot}$ at $z=0$. The model comparison in Section \ref{sec:model_comparison} uses the L1\_m9 simulations, all other parts of this work are based of the L2p8\_m9 simulation.

We remove all star forming gas particles from our analysis, because their temperature reflects the imposed  entropy floor and represents the average of the unresolved multiphase interstellar medium, in almost all cases they are anyway below $\mathrm{10^5 ~ K}$. Particles heated directly by an AGN in the last $15~\mathrm{Myr}$ are also removed. Because the AGN feedback model is likely not realistic on small scales, recently heated particles may have unrealistic temperatures and densities. Both the removal of star forming and recently heated gas particles have a negligible effect on the results. To compare with observations, we furthermore only consider particles which could reasonably emit non-negligible amounts of X-ray photons by imposing a temperature floor of $10^5 ~\mathrm{K}$. Without such a temperature floor, mass- and volume-weighted thermodynamic quantities would be less informative for our purposes. 

\begin{table*}
	\centering
	\caption{The observations to which the FLAMINGO data is compared. From left to right the columns list: Reference, sample size, median halo mass ($M_{\rm 500c}$), and the redshift of the observed objects.}
	\label{tab:example_table}
	\begin{tabular}{lrll} % four columns, alignment for each
		\hline
		Reference & sample size & $M_{\rm 500c} (\mathrm{M_\odot})$ & redshift \\
		\hline
		\citet[][]{Croston2008} & 31 &  $10^{14.1} - 10^{15.0}$ (median $10^{14.49}$) & $<0.2$  \\

  		\citet[][]{Pratt2009} & 31 &  $10^{14} - 10^{15}$ & 0.08-0.15  \\

		\citet[][]{Sun2009} & 43 &  $10^{13} - 10^{14}$ & $<0.12$ \\	
  
  		\citet[][]{Arnaud2010} & 33 &  $10^{14} - 10^{15}$ (median $10^{14.37}$) & $<0.2$  \\

  		\citet[][]{Sun2011} & 43 &  $10^{13} - 10^{14}$ & $<0.12$  \\

        \citet[][]{Sayers2013} & 45 & $10^{14.5} - 10^{15.4}$ (median $10^{14.95}$) & 0.15-0.89  \\

		\citet[][]{McDonald2014} & 80 &  $>10^{14.5}$ (median $10^{14.75}$) & $[0.3 - 0.6], [0.6 - 1.2]$   \\

  		\citet[][]{McDonald2017} & 147 &  $10^{14.0} - 10^{15.3}$ & $0.0 - 1.9$   \\

        \citet[][]{Lovisari2015} & 23 &  $10^{13} - 10^{14}$ & $<0.035$   \\

        \citet[][]{Planck2016} & 439 & $10^{14} - 10^{15}$ & $<0.25$ \\

        \citet[][]{Bulbul2019} & 59 & $10^{14.4} - 10^{15.1}$ & $0.2-1.5$ \\
  
		\citet[][]{Gaspari2019} & 85 & $10^{12} - 10^{14}$ & $<0.04$   \\

		\citet[][]{Ghirardini2019} & 12 &  $10^{14.5} - 10^{15}$ (median $10^{14.79}$) & $<0.1$  \\

		\citet[][]{Lovisari2020} & 120 &  $10^{14.3} - 10^{15.2}$ & $\sim 0.2$   \\
  
		\citet[][]{Migkas2020} & 313 &  $\rm >5 \times 10^{12} \; \mathrm{erg ~s^{-1}}$ & <0.3 (median 0.075)  \\
  
		\citet[][]{Ghizzardi2021} & 12 &  $10^{14.5} - 10^{15.0}$ (median $10^{14.79}$) & $<0.1$  \\	
        \hline
	\end{tabular} \label{table:observations}
\end{table*}

\subsection{Thermodynamic profiles}

To create radial profiles, we consider all particles eligible according to the above criteria. These are assigned to 30 logarithmically spaced radial bins between $\mathrm{0.01 ~ R_{\rm 500c}}$ and $\mathrm{3 ~R_{\rm 500c}}$. For each radial bin, we compute either the volume-, X-ray- or mass-weighted quantity of interest,
\begin{align}
    Q_V &= \frac{\sum_i q_i \times V_i}{\sum_i V_i} \, , \label{eq:volume_weight}\\
    Q_X &= \frac{\sum_i q_i \times L_{X,i}}{\sum_i L_{X,i}} \, , \label{eq:xray_weight} \\
    Q_M &=  \frac{\sum_i q_i \times m_i}{\sum_i m_i} \, , \label{eq:mass_weight} 
\end{align}
where $q_i$ is the value of the quantity for particle $i$, $m_i$ is the particle mass, $V_i$ its SPH volume ($V_i = m_i / \rho_i$) and $L_{X,i}$ its X-ray luminosity in the ROSAT band ($\mathrm{0.5-2.0~keV}$). The summation runs over all particles contained within the radial bin. All our profiles are 3D profiles, we have checked the effect of projection in Appendix \ref{sec:appendix_projection} and find it to be non-negligible. However, lacking a method to undo the deprojection in observational data, our work is limited to 3D space. When presenting the radial profiles, we ensure that at least 16\% of haloes have at least 1 particle in each bin, to ensure a proper representation of the $1\sigma$ error. This leads to different inner stopping radii for different mass haloes.

\subsubsection{Normalisation}
\paragraph*{Density}
The density profile is computed from the simulation using the free electron number density of each particle. The electron number density is computed internally in FLAMINGO by interpolating the tables from \citet[][]{ploeckinger2020}, which take into account the ionisation state given the temperature-, density-, metallicity- and redshift-dependent UV and X-ray background and interstellar radiation field, for each particle.

\paragraph*{Pressure}
The hot gas electron pressure is defined as
\begin{equation}
    P_{\rm e} = n_{\rm e} k_{\rm B} T \, ,
\end{equation}
where $n_{\rm e}$ is the free electron number density, $T$ the temperature of a particle in our simulation, and $k_{\rm B}$ the Boltzmann constant. The pressure is normalised relative to the virial equilibrium expectation $P_{500}$ (see Appendix~\ref{pressure_derivation} for a derivation)
\begin{equation}
    P_{500} = \frac{3}{8\pi} \left( \frac{500 G^{-1/4}}{2} H(z)^2 \right)^{4/3} f_{\rm B} \frac{\mu}{\mu_{\rm e}} M_{\rm 500c}^{2/3} \, ,
\end{equation}
where $G$ is the gravitational constant, and $H(z)$ the Hubble parameter. We use $f_{\rm B} = \Omega_{\rm b} / \Omega_{\rm m} = 0.159$ the cosmological baryon fraction in FLAMINGO, $\mu = 0.59$ the mean particle mass, and $\mu_{\rm e} = 1.14$ the mean particle mass per free electron.

\paragraph*{Temperature}
We normalise the temperature profiles to $T_{500}$ (see Appendix \ref{temperature_derivation} for a derivation)
\begin{equation} \label{eq:T500}
    T_{500} = \frac{\mu m_{\rm p}}{2 k_{\rm B}} \left( \frac{500 G^2}{2} \right)^{1/3} M_{\rm 500c}^{2/3} H(z)^{2/3} \, ,
\end{equation}
with $m_{\rm p}$ the proton mass.

\paragraph*{Entropy }
The entropy is defined as
\begin{equation}\label{eq:entropy_def}
    K = \frac{k_{\rm B} T}{n_{\rm e}^{2/3}} \, ,
\end{equation} 
where both the temperature and electron number densities are taken from the particles in our simulation. This is normalised relative to $K_{500}$ (see Appendix \ref{entropy_derivation} for a derivation)
\begin{equation}
    K_{500} = \frac{\left( \mu^5 m_{\rm p}^5 G^4 \right)^{1/3}}{5 \left( \frac{3}{\pi} \right)^{2/3} f_{\rm B}^{2/3}} M_{\rm 500c}^{2/3} H(z)^{-2/3} \, .
\end{equation}

\paragraph*{Metallicity}
We define the metallicity as the abundance of iron relative to hydrogen in solar units,
\begin{equation}
    \left[ \mathrm{Fe/H} \right] = \log_{\rm 10} \left(\frac{X_{\mathrm{Fe}}}{X_{\mathrm{H}}} / \frac{X_{\mathrm{Fe}, \odot}}{X_{\mathrm{H}, \odot}} \right) \, ,
\end{equation}
where $X_{\mathrm{Fe}}$ and $X_{\mathrm{H}}$ are the mass fractions of respectively iron and hydrogen.
We use the solar abundance ratio from  \citet{Asplund2009}, $X_{\mathrm{Fe}, \odot} / X_{\mathrm{H}, \odot} = 10^{7.5} / 10^{12}$.

\begin{figure*}
	\includegraphics[width=\linewidth]{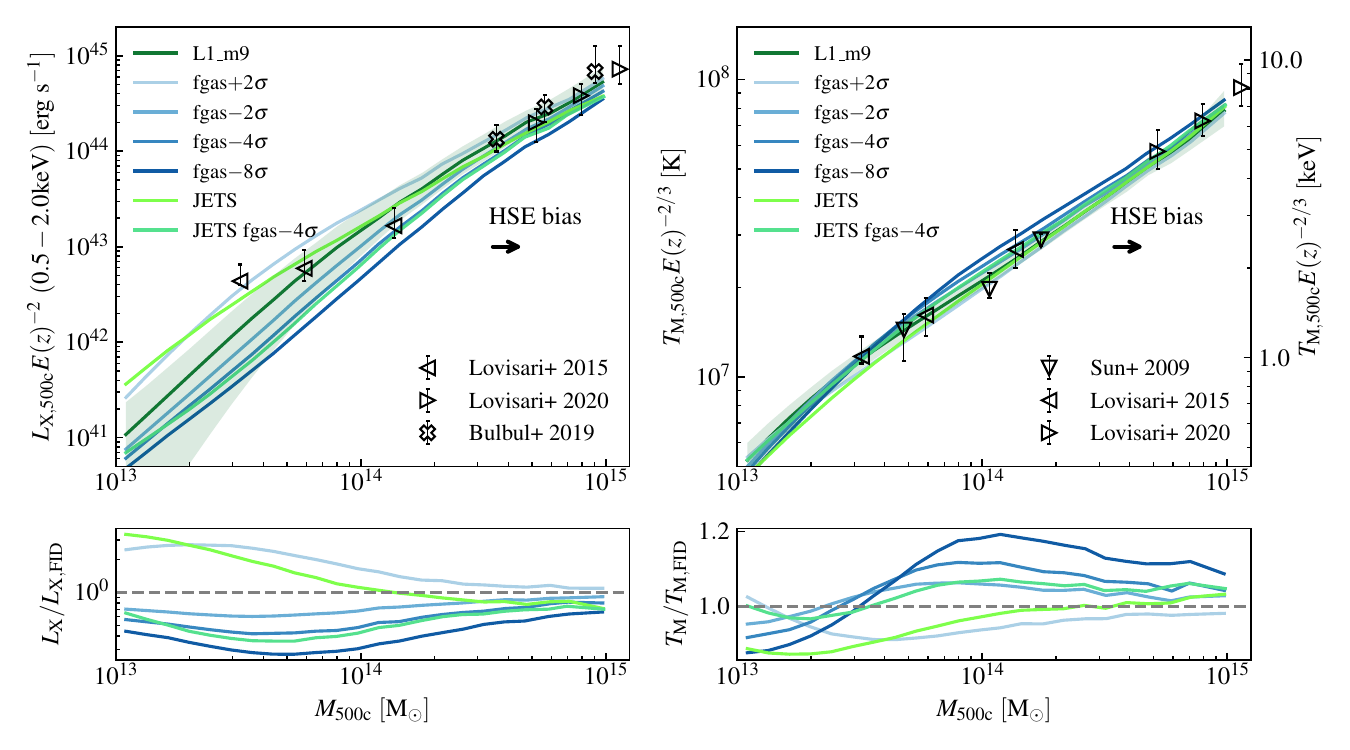}
    \caption{Variation in the cluster scaling relations for haloes with $M_{500\mathrm{c}} > 10^{13} ~\mathrm{M_\odot}$, for the different FLAMINGO models run at the fiducial numerical resolution. Lower panels show the ratio for the different variations with the fiducial (L1\_m9) model. The lines are the median relations between halo mass and X-ray luminosity (left), and between halo mass and temperature (right). Luminosities, temperatures and masses are measured within $R_{\rm 500c}$. Comparisons are made with X-ray data from \citet[][$0.08 < z < 0.15$]{Pratt2009}, \citet[][$z < 0.035$]{Lovisari2015}, \citet[][$0.059<z<0.546$]{Lovisari2020}, \citet[][$0.2<z<0.66$]{Bulbul2019}, \citet[][$z < 0.04$]{Gaspari2019}, and \citet[][$z < 0.3$]{Migkas2020}. The shaded region (L1\_m9 only) and the observational error bars indicate the 16th and 84th percentiles of the sample.}
    \label{fig:scaling_relations}
\end{figure*}

\section{Effect of model variations on the scaling relations} \label{sec:scaling_relations_models}
One of the primary purposes of FLAMINGO is the study of galaxy groups and clusters. Even though the simulations have been calibrated to match observed gas fractions at $\mathrm{R_{\rm 500c}}$ for clusters at $z \approx 0.1 - 0.3$ and $M_\text{500c} = 10^{13.5} - 10^{14.36}~\text{M}_\odot$, this does not guarantee the reproduction of cluster scaling relations, as those depend on density, temperature, and metallicity profiles, as well as on the clumpiness and multiphase nature of the ICM. Furthermore, calibration does not include all masses, and is only performed at low redshift, which also makes the scaling relations a test for the predictive power of the simulations. \citet{schaye2023} already showed that at $z=0$ the different resolutions match various observed scaling relations well across more than two decades in mass. Here we study the model dependence of the same scaling relations. 

Figure \ref{fig:scaling_relations} shows the median X-ray luminosity -- halo mass (left), and temperature -- halo mass (right) relations for all groups and clusters in the different models in a ($1~\mathrm{Gpc}$)$^3$ volume at the fiducial resolution. Temperatures are mass-weighted and include all particles with $T>10^5 ~\mathrm{K}$, we have compared with emission-weighted temperatures and found the differences to be within 10\% for these global properties. The small horizontal arrows indicate the systematic shift applied to observational data with hydrostatic-equilibrium inferred masses, which corresponds to the value found for the hydrostatic mass bias (0.743) during the calibration in \citet{Kugel2023}, which agrees with their assumed priors based on the observations of \citet{Hoekstra2015} and \citet{Eckert2016}. We neglect any corrections on the observed quantities due to the hydrostatic bias as these are negligible (e.g. the X-Ray luminosity will be dominated by radii much smaller than $R_{\rm 500c}$)

The observed X-ray luminosities have been shifted to the 0.5-2.0 keV band using PIMMS\footnote{\url{https://heasarc.gsfc.nasa.gov/docs/software/tools/pimms.html}} \citep{Mukai1993}. Observational data was grouped into a limited number of bins per data set, where the error bars show the scatter between individual objects in that bin.

The different models show a consistent offset between models with different gas fractions, which is most pronounced in the luminosity -- mass relation. Apart from the offset, the slope of the scaling relation varies with the model variations. The normalisation and slope, measured by fitting a single power law between $10^{13.5}-10^{15} ~\mathrm{M_{\odot}}$ are tabulated in Table \ref{tab:slope_and_norm}, according to
\begin{equation} \label{eq:powerlaw_fit}
    \mathcal{Y} = \alpha \times \log_{10} (m_{\rm 500c}) + \beta \, ,
\end{equation}
with $\mathcal{Y}$ either the logarithmic luminosity or temperature, $\alpha$ the slope and $\beta$ the normalisation.

As can be seen from Fig. \ref{fig:scaling_relations} the temperature -- mass relation shows little variation in the slope and normalisation above $10^{13.5} ~\mathrm{M_{\odot}}$, but the luminosity -- mass relation varies by 40\% in slope and by and order of magnitude in normalisation. The offset can be explained by the different gas fractions, with less gas producing less X-ray emission. The change in slope can be understood in the light of less massive objects having larger fractional changes in their gas fractions between the model variations. The paucity of gas, and hence X-ray luminosity, at lower masses, combined with the relatively small change at higher masses, yields a change in slope as seen in these relations.

\begin{table}
    \centering
    \caption{Slope ($\alpha$) and normalisation ($\beta$) of the scaling relations for simulations calibrated to different gas fractions (see eq. \ref{eq:powerlaw_fit}).}
    \begin{tabular}{l|llll}
         Model & $\alpha$ -- M-T  & $\beta$ -- M-T & $\alpha$ -- M-L & $\beta$ -- M-L  \\ \hline
         fgas$+2\sigma$ & 0.58 & 7.01 & 1.47 & 42.62 \\
         L1\textunderscore m9 & 0.55 & 7.05 & 1.75 & 42.24\\
		 fgas$-2\sigma$ & 0.55 & 7.07 & 1.88 & 42.01 \\
		 fgas$-4\sigma$ & 0.55 & 7.09 & 1.97 & 41.84 \\
		 fgas$-8\sigma$ & 0.57 & 7.09 & 2.04 & 41.63\\
         Jet & 0.59 & 7.01 & 1.45 & 42.48 \\
         Jet\_fgas$-4\sigma$ & 0.57 & 7.06 & 2.00 & 41.77\\
    \end{tabular}
    \label{tab:slope_and_norm}
\end{table}

It should be noted that, except for the most extreme model variations, the medians of all models are within $1\sigma$ of each other for both scaling relations across the entire mass range.

We have verified that using core-excised quantities (i.e.\ removing radii $r < 0.15~R_{\rm 500c}$) or spectroscopic-like temperatures \citep{Mazzotta2004, Vikhlinin2006} does not significantly change the scaling relations, nor their agreement with data. In summary, the simulated cluster scaling relations are numerically converged and agree well with observations, the FLAMINGO models provide observationally motivated variations which are measurably different in these integrated quantities.

\section{Thermodynamic profiles}\label{sec:profiles}
In the previous section we have shown that the scaling relations obtained from FLAMINGO clusters match observational data, and do so across a wide range of halo masses. Scaling relations are, however, by their very nature limited to comparing global properties of haloes. For many observables, global properties can be dominated by one region of the halo, such as the outer halo for mass-weighted temperatures, whereas X-ray properties tend to be sensitive to the core. There are scenarios where clusters fall on the scaling relations, but have unrealistic radial profiles. This section compares the density, temperature, metallicity, entropy, and pressure profiles from FLAMINGO with observational data. 

As described in Section \ref{sec:halo_selection}, we exclude star forming particles, particles that have recently been directly heated by AGN feedback, and cool gas ($T < 10^5 ~ \mathrm{K}$). We find that the exclusion of these three categories of particles has a very small impact on most quantities. Only the innermost part of the mass- and volume-weighted temperature profile is significantly affected when cool particles are included. 

Using the above selection criteria, we compute the profiles for all $M_{500\mathrm{c}}\rm > 10^{13}~ \mathrm{M_{\odot}}$ haloes. This is done using 30 equally spaced logarithmic bins between $0.01~R_{\rm 500c}$ and $3~R_{\rm 500c}$. In each bin we compute the volume-, emission-, or mass-weighted average of each physical quantity. Volume weighting is performed using the SPH volume of each particle (eq.~\ref{eq:volume_weight}). Emission- and mass-weighting are performed analogously, replacing the SPH volume by either the $\mathrm{[0.5-2.0] ~ keV}$ X-ray luminosity (eq.~\ref{eq:xray_weight}), or the mass of each particle (eq.~\ref{eq:mass_weight}). Using spectroscopic-like temperatures \citep{Mazzotta2004, Vikhlinin2006} results in negligible changes to the temperature profile over the entire radial range compared to X-ray luminosity weighting, hence we choose not to show it.

The smallest radii at which we show the median profile is determined by the radius at which at least 16\% of the haloes no longer have particles in the radial bin, this is the point where the $1\sigma$ error can no longer be reliably determined. This results in mass-, and resolution-dependent cut-off radii. For higher mass objects, we can follow the curves to smaller $r/R_{\rm 500c}$ simply because those objects have a higher density. 

The observations with which we compare our simulated cluster sample are listed in Table \ref{table:observations}, where we tabulate the sample size, mass range, median mass, and the redshift range of the observed sample. All observed normalised radii ($x = r / R_{\rm 500c}$) have been multiplied to account for the hydrostatic mass bias $x_{\rm hs} / x = (0.743)^{1/3}$ obtained during the calibration of FLAMINGO, as well as their normalised observed values. The metallicities found by \citet{Ghizzardi2021} have been corrected to the solar iron abundance in \citet{Asplund2009}. All datasets have been re-scaled to the FLAMINGO baryon fraction ($f_{\rm B} = 0.159$), mean particle mass ($\mu=0.59$) and mean particle mass per free electron ($\mu_{\rm e}=1.14$). 

To interpret the simulated thermodynamic profiles, and to compare them with observations, we first have to understand the impact of different mass selection effects and weighting choices. With this in mind, we will first study the mass dependence of simulated thermodynamic profiles in Section \ref{sec:mass_dependence}, then consider the different choices for weighting particles in Section \ref{sec:weighting_dependence}, and finally the difference between cool-core and non-cool-core clusters in Section \ref{sec:coolcore_dependence}. These three topics will set the stage for the FLAMINGO model comparison, which we show in Section \ref{sec:model_comparison}.

\begin{figure}
    \centering
    \includegraphics[width=1\linewidth]{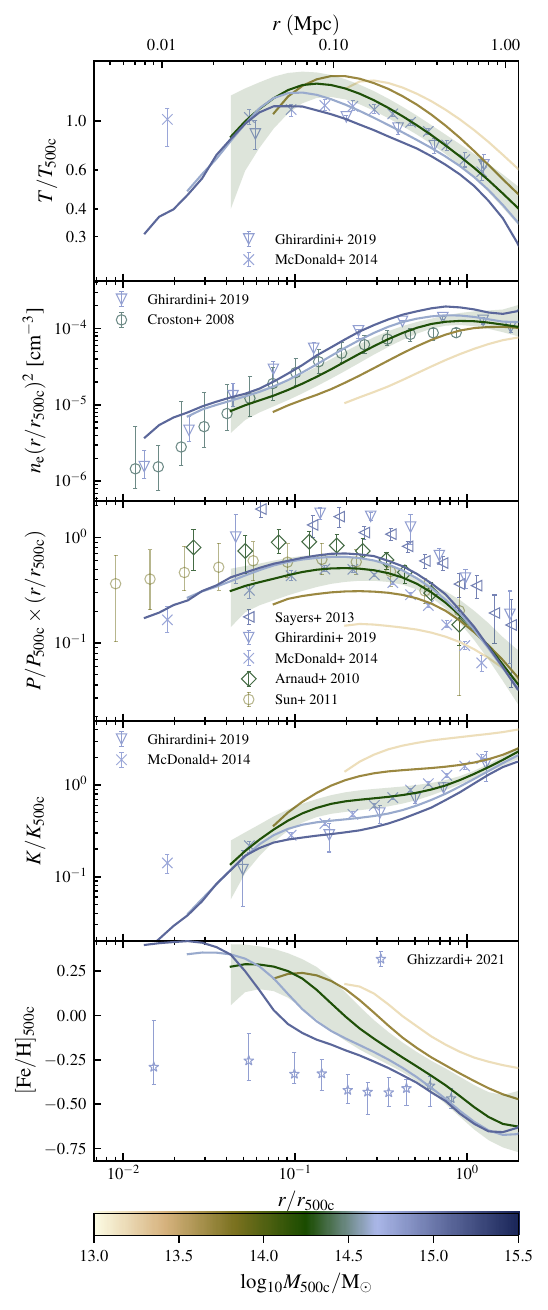}
    \caption{From top to bottom, the different panels show the median temperature, density, pressure, entropy, and metallicity profile for galaxy groups and clusters in the fiducial model (L1\_m9) of the FLAMINGO simulations at $z = 0$. All quantities are X-ray-luminosity-weighted. The color scale encodes halo mass ($M_{\rm 500c}$), as indicated by the color bar. Different lines are for different 0.5 dex mass bins. Data points correspond to observational data sets listed in Table \ref{table:observations}, and have a colour corresponding to their median mass. Error bars indicate the 16th and 84th percentiles when available, or the observational $1\sigma$ error. The shaded region indicates the 16th-84th percentiles for the $M_\text{500c} = 10^{14.0}-10^{14.5}\,\text{M}_\odot$ mass bin.}
    \label{fig:mass_dependence}
\end{figure}

\subsection{Mass dependence} \label{sec:mass_dependence}
The temperature, pressure, and entropy profiles in this work are normalised by the average value of each quantity expected within $R_{\rm 500c}$ assuming an isothermal sphere in hydrostatic equilibrium and the virial relations ($T_{\rm 500c}$, $P_{\rm 500c}$, and $K_{\rm 500c}$). This normalisation is naturally mass and redshift dependent as the total gravitational force exerted by a virialized structure depends on its mass, and the critical density evolves with redshift. In order to compare the profiles of different mass haloes, we use the normalised radius ($r / R_{\rm 500c}$). If all objects, of all masses, would be spherically symmetric, self-similar, virialised structures in hydrostatic equilibrium, then the normalized thermodynamic profiles would be mass independent, assuming that there are no non-gravititional contributions to the heating, cooling or gas kinetics. In reality this is most likely not the case, as the halo formation time and merger rate, and hence the departure from sphericity and equilibrium, are mass dependent, as well as the normalised radius out to which galaxy baryonic processes can have a discernible and significant impact. In addition, halo concentrations and stellar-to-halo mass ratios vary systematically with halo mass.

We find that the normalised profiles still show a strong mass dependence. Since observational campaigns tend to target, and are thus biased towards the higher-mass objects, and since mass estimation from observations is fraught with uncertainties, the mass dependence is important to keep in mind when comparing profiles from simulated objects with observations. 

Figure \ref{fig:mass_dependence} shows five $z=0$ median thermodynamic profiles (temperature, density, pressure, entropy, and metallicity) in 5 different mass bins (each 0.5 dex wide), stretching from $M_{\rm 500c} = \mathrm{10^{13}~ M_{\odot}}$ to $\mathrm{10^{15.5}~ M_{\odot}}$. The 16$^{\rm th}$-84$^{\rm th}$ percentile are shown for the $\mathrm{10^{14}~ M_{\odot}}$ to $\mathrm{10^{14.5}~ M_{\odot}}$ mass bin, but are comparable in both shape and magnitude for all other masses. The observations we compare with are coloured by their median mass in an identical fashion to the simulations. All profiles are clearly mass dependent, despite the applied normalizations. From one mass bin to the next, the magnitude of the profiles at fixed radius changes by $\approx 0.1-0.3~\mathrm{dex}$. The normalized temperature, normalized entropy and metallicity decrease with mass, whereas the density and normalized pressure increase. The peak of the temperature profile moves to smaller normalized radii when the mass increases. For the entropy profile, we observe that the flat part of the profile shifts to smaller normalized radii for higher-mass objects. Flattening of the entropy profile compared to observations has previously been discussed in detail by \citet{Altamura2023}, who found it to be common in simulations, and stronger than our simulations, but leave open the exact origin of the phenomenon.

Compared to recent work, we find a similar mass dependence to MilleniumTNG \citep{Pakmor2023}, and are in broad agreement with TNG-Clusters \citep{Lehle2023}. Compared to those works, FLAMINGO clusters show a strong drop in the temperature in the core, as well as a drop in the core entropy, as opposed to flat profiles. In Section \ref{sec:coolcore_dependence} we will show that this could be due to a different cool-core fraction. The mass dependence is also similar to what is seen in clusters from The Three Hundred project \citep{Li2023}, when taking into account that profiles evolve self-similarly as we will show in Section \ref{sec:redshift}.

Except for the metallicity, which is $\approx 0.3~\mathrm{dex}$ higher than observed, our simulated clusters follow the observed relations down to small radii, when comparing with observations which have a similar median mass. The metallicity is also high compared to previous simulations which tend to produce less iron in the core compared to observations \citep{Vogelsberger2018, Pearce2021}.

The extrapolations of the temperature and entropy profiles to smaller radii than sampled by the simulation do not seem to match the observations by \citet{Sun2009} ($M_{\rm 500c} \approx 10^{13.5} ~\mathrm{M_\odot}$, green-brown line) and  \citet{McDonald2014}\footnote{The \citet{McDonald2014} data shown is for their $z = 0.3 - 0.6$ bin, but we will show in Section \ref{sec:redshift} that the normalised profiles hardly evolve, making this an appropriate comparison even at $z=0$.} ($M_{\rm 500c} \approx 10^{14.75} ~\mathrm{M_\odot}$, light-blue line). However, we will show that at the smallest radii the profiles of cool-core and non-cool-core clusters diverge, with the non-cool-core objects agreeing with data. 

For all masses, the simulated pressure profiles slightly undershoot the observed pressures. In addition, the sloped are slightly too shallow, underprediciting the pressure in the centers and overpredicting it in the outskirts.

The consistent discrepancy between the predicted and observed metallicity could be explained by the assumed nucleosynthetic yields being too high, or by the total stellar masses being too high in the simulation. Figure 11 of \citet{schaye2023} compares the cluster stellar masses to observations. For $M_{\rm 500c} > 10^{14}~\mathrm{M_{\odot}}$ the total stellar masses are indeed too high, however, if only the stellar mass within $50~\mathrm{kpc}$ apertures is included, then the stellar masses are too low, which implies that the comparison is sensitive to the difficult to observe extended low surface brightness stellar envelopes of galaxies.

For the different panels, where different observational data sets at the same mass are available, the difference between consecutive mass bins is of roughly the same magnitude as the difference between those data sets (with the exception of the metallicity profiles). This implies that any differences seen between the simulated profiles and observations, could potentially be explained by uncertainties in the observations.

\subsection{Weighting scheme} \label{sec:weighting_dependence}
\begin{figure}
    \centering
    \includegraphics[width=1\linewidth]{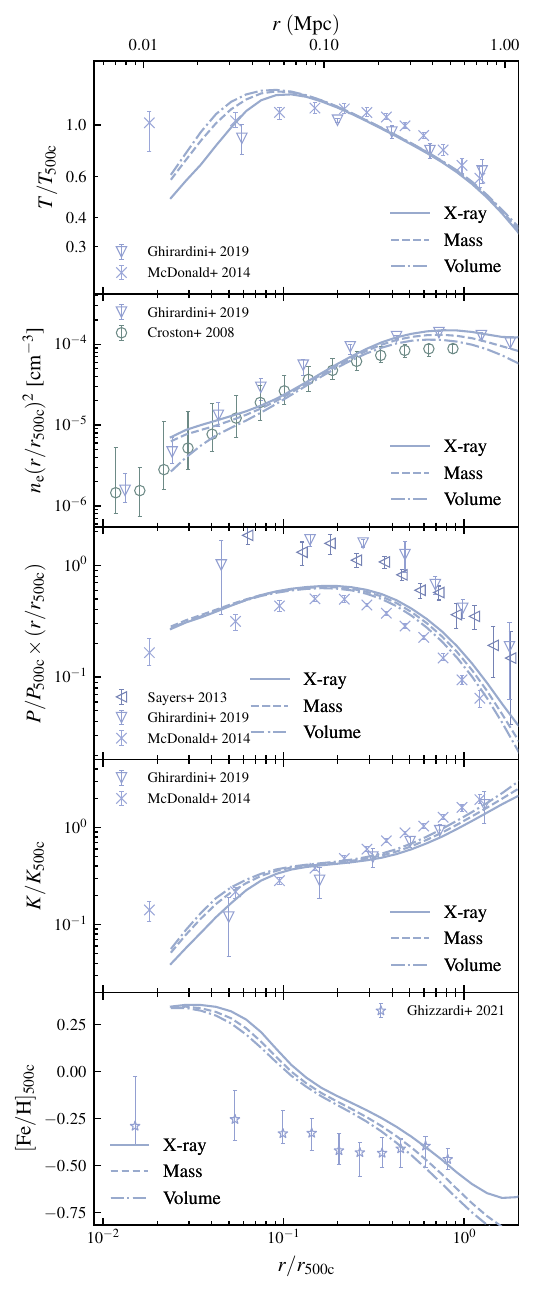}
    \caption{Similar to Fig \ref{fig:mass_dependence}, but only showing the fiducial model (L2p8\_m9) and massive clusters ($M_{\rm 500c} = 10^{14.5}-10^{15.0} {\rm M}_{\odot}$), using 3 different methods to weigh the particles when constructing the profiles. X-ray weighting is indicated by the solid line, mass weighting by the dashed line, and volume weighting by the dash-dotted line.}
    \label{fig:weighting}
\end{figure}
When calculating a cluster thermodynamic profile from simulations, an important choice is how to weigh the contribution of individual particles. The arguably most intuitive method, to weigh the contribution of each particle by the fractional volume it occupies in a spherical shell (see eq.~\ref{eq:volume_weight}), is not necessarily representative of what would be measured observationally. Because thermodynamic profiles are inferred from X-ray observations, X-ray bright gas will dominate the inference. The comparison with volume-weighted profiles from simulations is only fair if all volume elements within a spherical shell have the same X-ray brightness. 

To test this assumption, and to better approximate observational inferences, we compare three different weighting schemes. Volume-weighting, mass-weighting (see eq.~\ref{eq:mass_weight}), and X-ray-weighting (see eq.~\ref{eq:xray_weight}) in which particles are weighted by their density- and temperature-dependent X-ray luminosity. We expect the latter, which was used in Figure \ref{fig:mass_dependence}, to be closest to what is measured observationally. As a caveat we note that observationally the density, temperature and metallicity are measured from the radially-binned spectrum, we plan to include this in future work producing realistic synthetic X-ray observations.

In Figure \ref{fig:weighting} we compare the resulting profiles for the fiducial model (L2p8\_m9) and objects with masses between $M_{\rm 500c} = 10^{14.5}$ and $10^{15.0}~ \mathrm{M_{\odot}}$ (a total of 12354 objects). The physical radius (top axis) is computed for the median mass of the sample. The differences between the different weighting schemes are generally small, and comparable to or smaller than the scatter between different observations. The scatter in the simulations is similar for all weighting schemes and comparable in magnitude to what is shown in Fig. \ref{fig:mass_dependence}.
The differences due to weighting are generally much smaller than an $0.5 ~ \mathrm{dex}$ change in mass, as seen in Figure \ref{fig:mass_dependence}. However, for the temperature, density, and entropy the weighting causes larger differences in the core ($r \ll R_{\rm 500c}$) than a $0.5 ~\mathrm{dex}$ change in mass. Compared to volume-weighting, for mass- and particularly for X-ray-weighting, the inner temperature and entropy are lower, the density and metallicity higher, and the pressure is unaffected. Differences in the temperature profile are limited to $r < 0.1~r_{500\mathrm{c}}$. For the density and entropy profiles the differences increase towards both small and large radii. The pressure and metallicity profiles show the weakest dependence on the weighting scheme and an increasing difference with radius.

These tendencies are in line with expectations. Since mass and X-ray weighting favour denser gas, they yield a higher density, and this denser gas tends to have a lower temperature, lower entropy and higher metallicities. The differences increase towards small radii, where the gas tends to be more multiphase. Because the different phases tend to be in pressure equilibrium, the pressure profiles are nearly the same for the different weighting schemes. The difference at larger radii seen for the density and entropy might be due to particles bound to satellites, which will have higher densities. The differences between applying different weights are within the scatter between different observational data sets. We note, however, that when discrepancies with the observational data exist, they tend to diminish when moving from volume to mass weighting, and even more so when changing to X-ray luminosity weighting.

\subsection{Cool cores}\label{sec:coolcore_dependence}
\begin{figure}
    \centering
    \includegraphics[width=\linewidth]{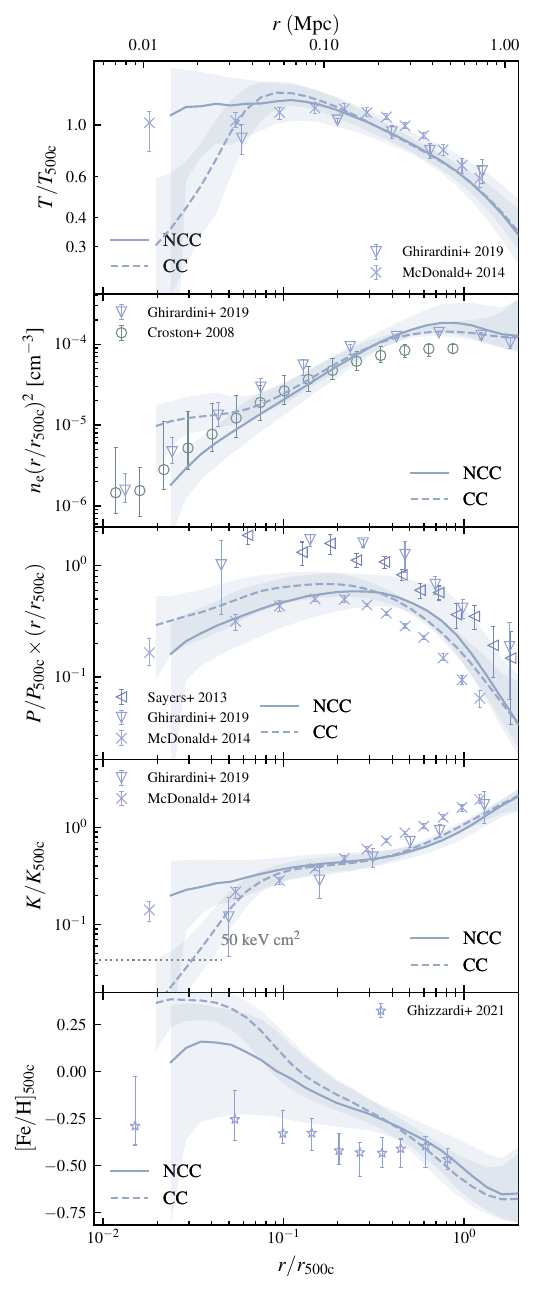}
    \caption{Similar to fig \ref{fig:mass_dependence}, but only showing the fiducial model (L2p8\_m9) for massive clusters ($M_{\rm 500c} = 10^{14.5}-10^{15.0} \mathrm{M_{\odot}}$), comparing cool-core (dashed) and non-cool-core samples (solid). The two samples show large differences in the cluster cores.}
    \label{fig:coolcore}
\end{figure}

We distinguish cool core (CC) and non-cool core (NCC) clusters in our simulations by measuring the radiative cooling time within $0.048 R_{\rm 500c}$. This is in line with \citet{hudson2010}, who compared 16 CC metrics and found this one to be the most distinguishing feature. We compute the cooling rate from the cooling tables used for FLAMINGO \citep{ploeckinger2020}. These are interpolated to the X-ray-luminosity-weighted temperature, density, and metallicity of particles within $0.048 R_{\rm 500c}$ (the median NCC cluster has 10 particles within this radius), where we make the same selection to exclude recently heated, star forming and cool particles as in the rest of this work. Using the radiative cooling rate $\Lambda$, we compute the cooling time
\begin{equation}\label{eq:cooling_time}
    t_{\rm cool} = \frac{6 n_{\rm e} k_{\rm B} T}{2 n_{\rm e}^2 \Lambda(n_{\rm e}, T, Z, z)} \, ,
\end{equation}
with $Z$ the gas metallicity. CC clusters are often defined as objects that have a central cooling time below a critical value, which tends to be set to a value between 1 and 5 $\mathrm{Gyr}$. To discern strong CC clusters from NCC clusters, we define the CC sample to be objects with $t_{\rm cool} < 1\;\mathrm{Gyr}$. Conversely, the NCC sample has $t_{\rm cool} > 5\;\mathrm{Gyr}$.

We compare the (emission-weighted) cluster gas profiles for CC and NCC clusters of mass between $M_{\rm 500c} = 10^{14.5}$ and $10^{15.0}~ \mathrm{M_{\odot}}$ in Fig.~\ref{fig:coolcore}\footnote{Note that for X-ray selected clusters the observational data tends to be skewed toward CC clusters \citep{Eckert2011}.}. In this mass bin, this selects 16\% of the total number of clusters as NCC (1988) and 22\% as CC (2785) (see also Fig.~\ref{fig:CCT_CDF}). We note that the sample median central cooling time of all objects is $2~\mathrm{Gyr}$, hence the total sample is more similar to CC than to NCC clusters. Since the cooling time decreases with density, we expect CC clusters to have denser cores. Cooling is also more efficient at sub-virial temperatures, and at higher metallicity, particularly below $1\;\mathrm{keV}$ where metal-line cooling becomes important. This gives rise to the expectation that CC clusters have cooler, higher metallicity gas in their cores compared to their NCC counterparts, something borne out in observations \citep[e.g.][]{DeGrandi2001, Lovisari2019}. We reproduce that trend in Fig.~\ref{fig:coolcore}, where the median relations for NCC and CC are shown. The differences between CC and NCC clusters are only large for $r \lesssim 0.1 R_{\rm 500c}$. The offset seen in the pressure profile at small radii could be explained if CC objects have more concentrated mass profile, which is borne out by the higher density in the core. We note that for temperature, density and entropy, the difference is the core surpasses the $1\sigma$ region shown as a shaded band.

CC clusters are sometimes defined as clusters with a low central ($r \lesssim 0.012~R_{\rm 500c}$) entropy, where the threshold tends to be in the range $30-50 \; \mathrm{keV cm^{2}}$ \citep{McDonald2013}. We see that our CC sample, which is selected based on the central cooling time, has a much lower central entropy compared to its NCC counterpart. The entropy at the smallest radius to which our profiles extend is below $50\; \mathrm{keV cm^2}$ for the CC sample, while it is more than double that value for the NCC sample. Our classification is thus in line with this alternative definition of CC clusters.  In Appendix \ref{sec:appendix_CC_criteria} we show that the cooling time criterion used clearly separates the CC and NCC populations in density, temperature and entropy, which shows that this is a robust way of identifying CC clusters in FLAMINGO.
\begin{figure}
    \centering
    \includegraphics[width=\linewidth]{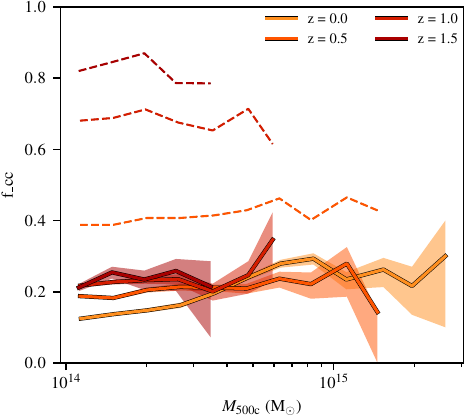}
    \caption{Fraction of objects classified as CC ($t_{\rm cool, (< 0.048\,R_{\rm 500c})} < 1 ~\mathrm{Gyr} \times E(z)^{-5/2}$) as a function of redshift. When correcting for self-similar evolution, there is no change in the CC fraction with redshift. The dashed lines show the result when using a non-evolving critical cooling time of $1 ~\mathrm{Gyr}$. Shaded regions indicate the $1\sigma$ bootstrap error on the CC fraction.}
    \label{fig:f_cc}
\end{figure}
\subsubsection{Evolution of the cool core fraction}
We study the evolution of the CC fraction of clusters over a wide mass range. As before, we define a CC cluster by its central ($r < 0.048~R_{\rm 500c}$) cooling time (eq.~\ref{eq:cooling_time}). At $z=0$ we used $1~\mathrm{Gyr}$ as the critical value for an object to be classified as a CC. The CC fraction we obtain is similar to that obtained by TNG-Cluster \citep{Lehle2023}, and slightly higher than IllustrisTNG \citep{Barnes2018}. Assuming the radiative cooling is dominated by bremsstrahlung, and that the virial temperature evolves self-similarly as $H(z)^{2/3}$ (eq.~\ref{eq:T500}), the cooling time of a cluster will evolve as $t_{\rm cool} \propto E(z)^{-5/2}$. Employing a self-similarly evolving critical value, Fig.~\ref{fig:f_cc} show that the CC fraction is almost constant with redshift. Only at the highest masses does it decrease slightly. The dashed lines show the result when not accounting for self-similar evolution, in that case a large fraction of objects is classified as CC at higher redshift. This strong evolution of the CC fraction is analogous to what has been found with a slightly different cooling time criterion by \citet{Barnes2018} for the IllustrisTNG simulation. However, it seems in conflict with observations which find an almost non-evolving CC fraction for a non-evolving cooling time criterion \citep{McDonald2017, Ruppin2021}.

\begin{figure*}
    \centering
	\includegraphics[width=.9\linewidth]{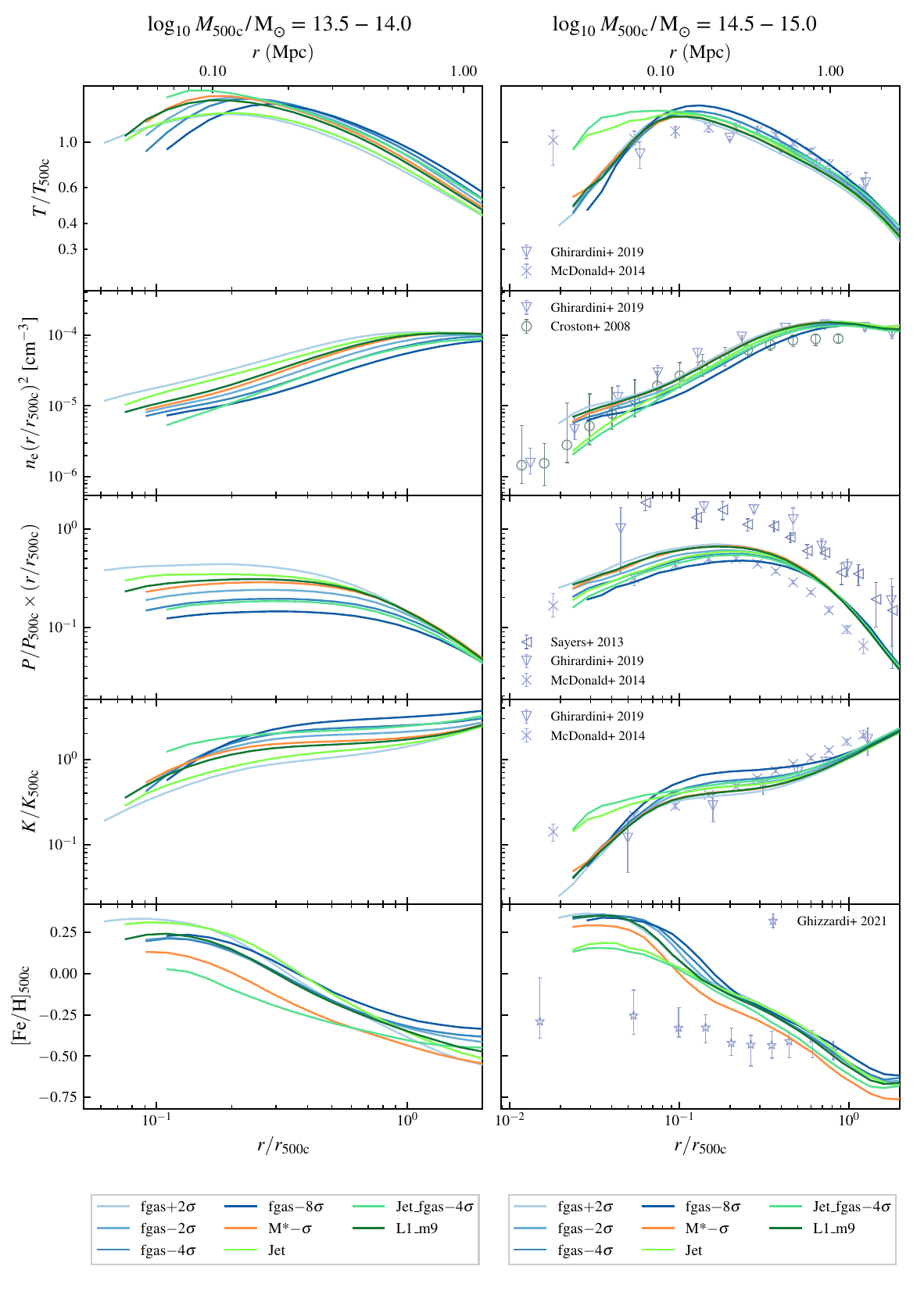}
    \caption{Similar to Fig.~\ref{fig:weighting}, but showing different FLAMINGO models in the same box size and for the same resolution as L1\_m9 for groups with mass $M_{\rm 500c} = 10^{13.5}-10^{14} ~\mathrm{M_\odot}$ in the left panel, and clusters with mass $M_{\rm 500c} = 10^{14.5}-10^{15} ~\mathrm{M_\odot}$ in the right panel. Except for the jet models and the most extreme gas fraction variation ($\mathrm{fgas-8\sigma}$), all model medians for clusters are within the 1$\sigma$ scatter of the fiducial model. The two jet models show distinctively different behaviour in the cores of clusters. For groups, the effect of model variations is much larger.}
    \label{fig:basic_profiles}
\end{figure*}

\subsection{Model comparison} \label{sec:model_comparison}
\begin{figure}
    \centering
    \includegraphics[width=\linewidth]{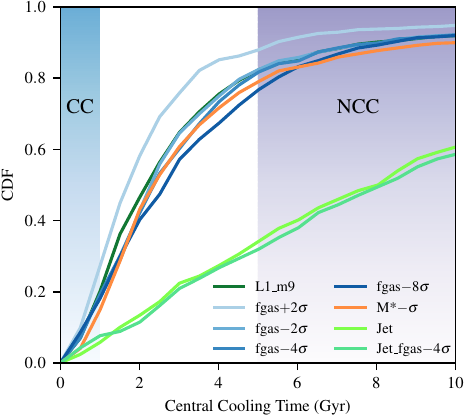}
    \caption{Cumulative distribution function of the central cooling times ($t_{\rm cool}(r < 0.048 \rm R_{\rm 500c}))$ of massive clusters ($M_{\rm 500c} = 10^{14.5}-10^{15} ~\mathrm{M_{\odot}}$) for all L1\_m9 model variations. The two jet models  have significantly greater central cooling times.}
    \label{fig:CCT_CDF}
\end{figure}

Using the knowledge from the previous three sections, we compare thermodynamic profiles for the different model variations in the FLAMINGO suite. We keep in mind the mass-dependence, which, depending on the quantity and radius (see Fig.~\ref{fig:mass_dependence}) can give a 20-100\% difference between consecutive mass-bins; the weighting dependence, which showed a strong radial dependence and can, at specific radii, change the result by 50\% (see Fig.~\ref{fig:weighting}); and the cool-core non-cool-core distinction, which is especially relevant for the innermost $0.1 R_{\rm 500c}$ of the thermodynamic profiles (see Fig.~\ref{fig:coolcore}).

Fig.~\ref{fig:basic_profiles} shows the different model variations for two different mass bins, large groups ($M_{\rm 500c} = 10^{13.5}-10^{14} ~\mathrm{M_\odot}$) in the left panel, and clusters ($M_{\rm 500c} = 10^{14.5}-10^{15} ~\mathrm{M_\odot}$) in the right panel. For all models, the scatter is comparable to what is shown in Fig. \ref{fig:mass_dependence}. The observational data we have selected is only available for massive objects, hence we do not show it for the lower mass bin.
First, we focus on the gas fraction model variations, which are calibrated to observed mass-dependent gas fractions shifted up or down by different numbers of observational $\sigma$ (see Table \ref{table:variations}). These are shown in different shades of blue in Fig.~\ref{fig:basic_profiles}. For $r \gtrsim 0.1~R_{\rm 500c}$, we see the expected trends: lower gas fractions correspond to lower densities, 
which implies a lower gas mass, and a lower gas pressure. The changes in density and pressure then determine the change in temperature which can go up or down depending on whether the density or pressure changes more. In FLAMINGO we find that the peak temperature decreases with increasing gas fraction. From the combination of a lower temperature and higher density with increasing gas fractions it follows that entropies are lower (see eq.~\ref{eq:entropy_def}). 

We furthermore find that the gas fraction variations show the largest differences at intermediate radii between $0.1-0.4 R_{\rm 500c}$, and converge at both larger and smaller radii (only visible for the cluster mass bin), except for the pressure profiles. The pressure profiles show an increasing difference between the models for decreasing radii, with lower gas fractions yielding up to 40\% lower pressures in the core of clusters, and a factor 4 lower pressure in the core of groups.

The lower gas fraction variations have a slightly stronger AGN feedback. The gas pushed out by the AGN is seemingly predominantly deposited between $0.1-0.4 R_{\rm 500c}$, as the difference between the models is largest there. The stronger AGN feedback deposits an increased amount of metals at these radii, explaining the divergence in metallicities in this radial range for clusters. At even larger radii, the models converge again. The correlation between temperature and gas fraction seen between $0.1-1 R_{\rm 500c}$ could also be due to the cumulative effect of the AGN. With the cosmic baryon fraction fixed between different gas fraction variations, a lower cluster gas fraction implies more gas has been expelled at early time. While clusters grow, this hot expelled gas will flow towards the cluster, naturally leading to higher temperatures in the outer regions. Though \citet{Mccarthy2011} found that the infalling gas has higher entropy not because it was heated at earlier times, but simply because it is taking the place of the low entropy material that was ejected.

The effect of a change in the cluster stellar fraction is seen in the stellar mass variation model, shown by the orange line. This model has a lower stellar mass, but the same gas fraction. Though almost identical to the fiducial model in temperature, density, pressure, and entropy, the decreased production of metals, due to a lower stellar fraction, yields a consistently lower metallicity at all radii. The lower metallicity for groups in the jet\_fgas$-4\sigma$ model can be explained by the fact that at this halo mass the stellar mass of objects in this model variation is even slightly below the stellar mass variation model ($M^* - \sigma$).

The two jet models show very different behaviour. Compared with the fiducial thermally driven AGN feedback, the kinetic jet feedback clearly expels more gas from the cluster core. This is seen in a depressed central density for clusters. The remaining gas in the core is heated to a significantly higher temperature compared to all thermal feedback models. 
The combination of a higher temperature and a lower density leads to a significantly higher entropy in the core, while the pressure remains similar. This is not true for groups, which can be explained by the gas fraction for the Jet model being similar to the fgas$+2\sigma$ variation in this mass range, and consequently, the Jet model is more similar to the increased gas fraction variation for groups.
The jet models also show a strong deficit of metals in the core of clusters compared to the fiducial model. We interpret this as a sign that our kinetic jet feedback model is more effective at expelling enriched gas from the cluster core. Previous work based on the OWLS simulations \citep{Schaye2010}, which use thermally-driven AGN feedback, has found that the profiles are mostly affected by ejection of gas from the high-$z$ progenitors of $z=0$ clusters \citep{Mccarthy2011}, which raises the interesting question of exactly when and where the feedback was actually injected. 
Though the relative lack of metals in the core is similar for the reduced gas fraction jet model ($\mathrm{Jet\_fgas-4\sigma}$), the overall metallicity is significantly lower for this model, especially at larger radii. We note that for this latter model, the stellar fraction is significantly lower \citep{schaye2023}, possibly due the expulsion of low-entropy gas limiting radiative cooling and thus reducing the overall central stellar production, and hence the metal production. However, we remind the reader that all variations have been calibrated to match the stellar mass function over the same range of masses.

The combination of distinctly higher temperatures, lower densities and higher entropies, is a typical signature for NCC clusters seen when comparing with CC clusters in observations \citep[e.g.][]{hudson2010}. In Fig.~\ref{fig:CCT_CDF} we compare the CC fraction of the jet models with all other models for clusters, finding that their fraction is indeed suppressed by more than 50\%. The different CC to NCC ratio in the jet sample could then possibly account for some of the difference we see in the temperature, density and entropy profiles. NCC clusters are also observed to have lower core metallicities \citep{Lovisari2019}, in line with the difference between the Jet models and the thermal AGN models in Fig.~\ref{fig:basic_profiles}.

\begin{figure}
    \centering
    \includegraphics[width=\linewidth]{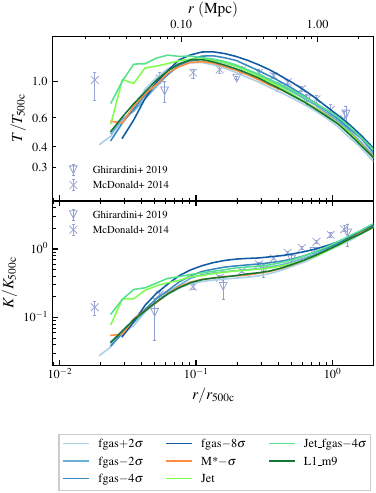}
    \caption{Temperature and entropy profiles for the central-cooling-time matched sample of massive clusters ($M_{\rm 500c} = 10^{14.5}-10^{15} ~\mathrm{M_{\odot}}$). After matching, the Jet and $\mathrm{Jet\_fgas-4\sigma}$ model models are closer to all other variations, but still discrepant.}
    \label{fig:CCT_matched_profiles}
\end{figure}

\subsubsection{Matching the cooling-time distributions}
Since the model variations have different strengths of their AGN and stellar feedback, they will also have different central cooling times. We have shown in Fig.~\ref{fig:coolcore} that the thermodynamic profiles of CC and NCC clusters differ, especially in the core region. The core region is also where we see a strong difference between the median profile for the jet models and all other variations, which begs the question of whether these different profiles give rise to different CC fractions between the model variations, and whether correcting for this would give consistent profiles between all models.

To study this effect, we first compute the cumulative distribution function (CDF) of the central-cooling times for clusters ($t_{\rm cool}(r < 0.048~R_{\rm 500c})$, eq.~\ref{eq:cooling_time}). We plot these distributions in Fig.~\ref{fig:CCT_CDF} which clearly shows that the jet models have much larger central cooling times compared to all other variations. The median profiles for the jet models will thus be much more NCC-like compared to the other variations.

To test whether this can explain the large difference seen in the cores for the median thermodynamic profiles of clusters, we create a matched sample of clusters across all model variations for the cluster mass bin ($M_{\rm 500c} = 10^{14.5}-10^{15} ~\mathrm{M_{\odot}}$), selecting for every object in the fiducial model (L1\_m9) an object in each variation that has a central cooling time differing by less than 10\% without replacement. If such a matching object cannot be found for every model variation, the halo is discarded. In this manner, we match $\approx 1/3$ of all haloes in the simulations. This matched sample should have nearly identical distributions of central cooling times, and we have checked that their CDFs are indeed nearly identical. For this matched sample of haloes, we recompute the median thermodynamic profiles for all variations. Fig.~\ref{fig:CCT_matched_profiles} shows these profiles for the temperature and entropy, for which the biggest difference with the two jet variations were seen. We see that after matching the cooling time distributions, the median profiles for the Jet and $\mathrm{Jet\_fgas-4\sigma}$ models are more similar to all other variations, but still show differences in the core. This suggests that the differences in Fig.~\ref{fig:basic_profiles} are only partly due to the different CC/NCC sample composition.

\begin{figure}
    \centering
    \includegraphics[width=\linewidth]{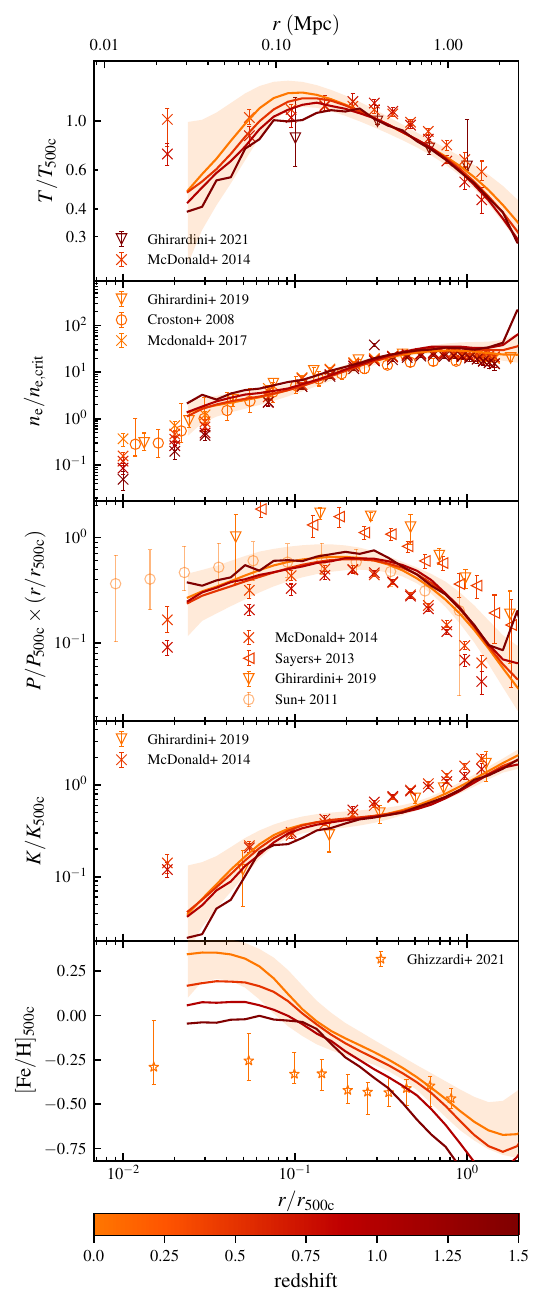}
    \caption{As figure \ref{fig:mass_dependence}, Cluster gas profiles for cluster with mass ($M_{\rm 500c} = 10^{14.5}-10^{15} ~\mathrm{M_\odot}$) at different redshifts indicated by the colour. Observations are coloured by their mean redshift. The physical radius (top axis) for $z=0$ objects is shown. Except for temperature and metallicity very little evolution is seen.}
    \label{fig:multiplot_redshift}
\end{figure}

\subsection{Redshift evolution at fixed mass} \label{sec:redshift}
We study the evolution of cluster gas profiles with redshift in Fig.~\ref{fig:multiplot_redshift} for the L2p8\_m9 simulation, selecting objects with the same mass ($M_{\rm 500c}$) at $z=0, 0.5, 1.0$, and $1.5$. Because haloes accrete mass between these redshifts, the most massive objects at $z=1.5$ may no longer be in the same mass bin at lower redshifts. However, by selecting the same mass range, we can compare the physical properties of haloes at a fixed mass over cosmic time. At $z=1.5$ there are 17 objects in this mass bin. At all redshifts, including $z=1.5$, the scatter is almost identical to the error band shown for $z=0$.

We find that the temperature, density, pressure, and entropy profiles show little to no evolution with redshift, when normalised to their redshift-dependent $X_{500}$ quantity, or the redshift-dependent critical density. This is expected since, although the Universe has expanded since $z=1.5$, leading to the gradual dilution of gas, the virial quantities $X_{\rm 500c}$ account for the effect of this expansion. It suggests that in FLAMINGO non-gravitational physics may impact (proto-)clusters mostly at high redshifts, as there is almost no deviation from self-similar evolution below $z=1.5$. In Appendix \ref{sec:appendix_scaling_relations} we show that scaling relations also evolve close to self-similarly.

Observations of high-redshift galaxy clusters find a deviation from self-similar expectation in the core ($r < 0.1 ~ R_{\rm 500c}$), with the cores evolving less since at least $z=1$ \citep{Ghirardini2021}. When using normalised quantities, this results in higher-redshift profiles being lower in the core compared to lower-redshifts, as the normalisation over-compensates for evolution. We see this most strongly in our temperature and entropy profiles, though the effect is not significant. Outside $0.1 ~ R_{\rm 500c}$ they are consistent with self-similarity, but inside that radius, the evolution is slightly slower. Unlike what is seen in observations \citep{McDonald2017, Ruppin2021}, the density and pressure profiles do not show deviations from self-similarity in the core.

The metallicity shows a clear evolution with redshift. Higher redshift clusters have a lower gas metallicity, particularly in the core. This is expected, since fewer stars have been formed by those earlier times, resulting also in fewer SNe Ia, which means that less metals have been injected into the ICM. The decrease in metallicity with redshift at large radii is consistent with what has previously been found in IllustrisTNG, C-EAGLE, and Magneticum \citep{Vogelsberger2018, Pearce2021, Angelinelli2023}. In contrast to those simulations, we find that the core metallicity also decreases with redshift.

\section{Conclusions} \label{sec:discussion} \label{sec:conclusion}
In this work we explored groups and clusters in the FLAMINGO cosmological hydrodynamical simulations, investigating the effect of mass selection, particle weighting, and cool-core selection, comparing their temperature, density, pressure, entropy, and iron abundance profiles to observations, and studied the evolution of cluster scaling relations and profiles. This has allowed us to conclude that, with the exception of the metallicity of the ICM, FLAMINGO clusters are a good match to objects in the real Universe, showing similar radial profiles and scaling relations.
In particular, we found that
\begin{enumerate}
    \item Even when normalised to virial expectations, thermodynamic profiles are sensitive to the mass of the selected objects. The normalised temperature, normalised entropy and iron abundance decrease with increasing mass, whereas the density and normalised pressure increase (Fig.~\ref{fig:mass_dependence}). In particular, the density profile shows a radius-independent offset of $\approx 0.2~\mathrm{dex}$ per $\mathrm{0.5~dex}$ in mass, which is consistent with the view that more massive objects originate from larger fluctuations in the background density field, yielding higher overall densities at all radii. Offsets in other thermodynamic profiles are of similar size but appear only at either small or large radii.
    \item Our fiducial weighting of particles when constructing the profiles is by X-ray luminosity. We find that using mass or volume weighting instead of X-ray weighting has a large, and non-trivial influence on the thermodynamic profiles (Fig.~\ref{fig:weighting}). While the temperature (pressure) profiles only show a difference of $\mathrm{0.3~dex}$ and $\mathrm{0.1~dex}$ in the innermost (outermost) regions, the density and entropy show a difference of $\mathrm{0.2~dex}$ at both large and small radii, with a decreased difference at intermediate radii.
    \item We define cool-core clusters using the central cooling-time, and show that in FLAMINGO they have lower core temperatures and entropies, but higher densities, pressures, and metallicities (Fig.~\ref{fig:coolcore}). This is in line with observational results, and implies that the sample composition is important to consider when comparing simulated clusters with observed clusters as the latter tend to have sample compositions biased towards more cool-core clusters. The cool-core fraction evolves strongly with redshift when using a non-evolving cool-core definition. When correcting for self-similar evolution, we find the cool-core fraction to be almost constant with redshift (Fig.~\ref{fig:f_cc}).
    \item The models using jet-like AGN feedback instead of our fiducial thermally-driven AGN feedback give thermodynamic profiles with qualitatively different shapes (Fig.~\ref{fig:basic_profiles}). Cluster cores are significantly hotter and less dense in these models, indicating that our kinetic jet feedback is effective at shaping the cluster cores.
    \item Except for the iron abundance, FLAMINGO galaxy clusters show little evolution in their thermodynamic profiles when normalised to their respective virial quantities (Fig.~\ref{fig:multiplot_redshift}). This indicates that massive galaxy clusters evolve self-similarly.
\end{enumerate}

We have shown that galaxy clusters from the FLAMINGO simulation evolve self-similarly, in line with expectations from observations. The thermodynamic profiles are sensitive to mass, even after normalisation to virial expectations, potentially due to the importance of non-gravitational physics for galaxy groups and clusters. For the cores of clusters, the weighting scheme chosen to measure physical quantities from simulations is important to consider. Finally, we have shown that the cool-core to non-cool-core composition of galaxy cluster samples should be matched to perform reliable comparisons of thermodynamic profiles.
In future work it would be useful to compare virtual observations to the data, which would allow for a fairer comparison and enable accounting for selection and projection effects.

\section*{Acknowledgements}
JB acknowledges support from research programme Athena 184.034.002 from the Dutch Research Council (NWO).
This work used the DiRAC@Durham facility managed by the Institute for Computational Cosmology on behalf of the STFC DiRAC HPC Facility (www.dirac.ac.uk). The equipment was funded by BEIS capital funding via STFC capital grants ST/K00042X/1, ST/P002293/1, ST/R002371/1 and ST/S002502/1, Durham University and STFC operations grant ST/R000832/1. DiRAC is part of the National e-Infrastructure. This project has received funding from the European Research Council (ERC) under the European Union’s Horizon 2020 research and innovation programme (grant agreement No 769130).

%%%%%%%%%%%%%%%%%%%%%%%%%%%%%%%%%%%%%%%%%%%%%%%%%%
\section*{Data Availability}
The data underlying the plots within this article are available on
reasonable request to the corresponding author. The FLAMINGO simulation data will eventually be made publicly available, though we note that the data volume (several petabytes) may prohibit us from simply placing the raw data on a server. In the meantime, people interested in using the simulations are encouraged to contact the corresponding author

%%%%%%%%%%%%%%%%%%%% REFERENCES %%%%%%%%%%%%%%%%%%

% The best way to enter references is to use BibTeX:

\bibliographystyle{mnras}
\bibliography{joey} % if your bibtex file is called example.bib

\begin{thebibliography}{}
\makeatletter
\relax
\def\mn@urlcharsother{\let\do\@makeother \do\$\do\&\do\#\do\^\do\_\do\%\do\~}
\def\mn@doi{\begingroup\mn@urlcharsother \@ifnextchar [ {\mn@doi@} {\mn@doi@[]}}
\def\mn@doi@[#1]#2{\def\@tempa{#1}\ifx\@tempa\@empty \href {http://dx.doi.org/#2} {doi:#2}\else \href {http://dx.doi.org/#2} {#1}\fi \endgroup}
\def\mn@eprint#1#2{\mn@eprint@#1:#2::\@nil}
\def\mn@eprint@arXiv#1{\href {http://arxiv.org/abs/#1} {{\tt arXiv:#1}}}
\def\mn@eprint@dblp#1{\href {http://dblp.uni-trier.de/rec/bibtex/#1.xml} {dblp:#1}}
\def\mn@eprint@#1:#2:#3:#4\@nil{\def\@tempa {#1}\def\@tempb {#2}\def\@tempc {#3}\ifx \@tempc \@empty \let \@tempc \@tempb \let \@tempb \@tempa \fi \ifx \@tempb \@empty \def\@tempb {arXiv}\fi \@ifundefined {mn@eprint@\@tempb}{\@tempb:\@tempc}{\expandafter \expandafter \csname mn@eprint@\@tempb\endcsname \expandafter{\@tempc}}}

\bibitem[\protect\citeauthoryear{{Abbott} et~al.,}{{Abbott} et~al.}{2022}]{Abbott2022}
{Abbott} T.~M.~C.,  et~al., 2022, \mn@doi [\prd] {10.1103/PhysRevD.105.023520}, \href {https://ui.adsabs.harvard.edu/abs/2022PhRvD.105b3520A} {105, 023520}

\bibitem[\protect\citeauthoryear{{Altamura}, {Kay}, {Bower}, {Schaller}, {Bah{\'e}}, {Schaye}, {Borrow}  \& {Towler}}{{Altamura} et~al.}{2023}]{Altamura2023}
{Altamura} E.,  {Kay} S.~T.,  {Bower} R.~G.,  {Schaller} M.,  {Bah{\'e}} Y.~M.,  {Schaye} J.,  {Borrow} J.,   {Towler} I.,  2023, \mn@doi [\mnras] {10.1093/mnras/stad342}, \href {https://ui.adsabs.harvard.edu/abs/2023MNRAS.520.3164A} {520, 3164}

\bibitem[\protect\citeauthoryear{{Angelinelli}, {Ettori}, {Dolag}, {Vazza}  \& {Ragagnin}}{{Angelinelli} et~al.}{2023}]{Angelinelli2023}
{Angelinelli} M.,  {Ettori} S.,  {Dolag} K.,  {Vazza} F.,   {Ragagnin} A.,  2023, \mn@doi [\aap] {10.1051/0004-6361/202245782}, \href {https://ui.adsabs.harvard.edu/abs/2023A&A...675A.188A} {675, A188}

\bibitem[\protect\citeauthoryear{{Arnaud}, {Pratt}, {Piffaretti}, {B{\"o}hringer}, {Croston}  \& {Pointecouteau}}{{Arnaud} et~al.}{2010}]{Arnaud2010}
{Arnaud} M.,  {Pratt} G.~W.,  {Piffaretti} R.,  {B{\"o}hringer} H.,  {Croston} J.~H.,   {Pointecouteau} E.,  2010, \mn@doi [\aap] {10.1051/0004-6361/200913416}, \href {https://ui.adsabs.harvard.edu/abs/2010A&A...517A..92A} {517, A92}

\bibitem[\protect\citeauthoryear{{Asplund}, {Grevesse}, {Sauval}  \& {Scott}}{{Asplund} et~al.}{2009}]{Asplund2009}
{Asplund} M.,  {Grevesse} N.,  {Sauval} A.~J.,   {Scott} P.,  2009, \mn@doi [\araa] {10.1146/annurev.astro.46.060407.145222}, \href {https://ui.adsabs.harvard.edu/abs/2009ARA&A..47..481A} {47, 481}

\bibitem[\protect\citeauthoryear{{Bah{\'e}} et~al.,}{{Bah{\'e}} et~al.}{2022}]{Bahe2022}
{Bah{\'e}} Y.~M.,  et~al., 2022, \mn@doi [\mnras] {10.1093/mnras/stac1339}, \href {https://ui.adsabs.harvard.edu/abs/2022MNRAS.516..167B} {516, 167}

\bibitem[\protect\citeauthoryear{{Barnes} et~al.,}{{Barnes} et~al.}{2017}]{Barnes2017}
{Barnes} D.~J.,  et~al., 2017, \mn@doi [\mnras] {10.1093/mnras/stx1647}, \href {https://ui.adsabs.harvard.edu/abs/2017MNRAS.471.1088B} {471, 1088}

\bibitem[\protect\citeauthoryear{{Barnes} et~al.,}{{Barnes} et~al.}{2018}]{Barnes2018}
{Barnes} D.~J.,  et~al., 2018, \mn@doi [\mnras] {10.1093/mnras/sty2078}, \href {https://ui.adsabs.harvard.edu/abs/2018MNRAS.481.1809B} {481, 1809}

\bibitem[\protect\citeauthoryear{{Barnes}, {Vogelsberger}, {Pearce}, {Pop}, {Kannan}, {Cao}, {Kay}  \& {Hernquist}}{{Barnes} et~al.}{2021}]{Barnes2021}
{Barnes} D.~J.,  {Vogelsberger} M.,  {Pearce} F.~A.,  {Pop} A.-R.,  {Kannan} R.,  {Cao} K.,  {Kay} S.~T.,   {Hernquist} L.,  2021, \mn@doi [\mnras] {10.1093/mnras/stab1276}, \href {https://ui.adsabs.harvard.edu/abs/2021MNRAS.506.2533B} {506, 2533}

\bibitem[\protect\citeauthoryear{{Bartalucci} et~al.,}{{Bartalucci} et~al.}{2023}]{Bartalucci2023}
{Bartalucci} I.,  et~al., 2023, \mn@doi [\aap] {10.1051/0004-6361/202346189}, \href {https://ui.adsabs.harvard.edu/abs/2023A&A...674A.179B} {674, A179}

\bibitem[\protect\citeauthoryear{{Biffi} et~al.,}{{Biffi} et~al.}{2016}]{Biffi2016}
{Biffi} V.,  et~al., 2016, \mn@doi [\apj] {10.3847/0004-637X/827/2/112}, \href {https://ui.adsabs.harvard.edu/abs/2016ApJ...827..112B} {827, 112}

\bibitem[\protect\citeauthoryear{{Bleem} et~al.,}{{Bleem} et~al.}{2015}]{Bleem2015}
{Bleem} L.~E.,  et~al., 2015, \mn@doi [\apjs] {10.1088/0067-0049/216/2/27}, \href {https://ui.adsabs.harvard.edu/abs/2015ApJS..216...27B} {216, 27}

\bibitem[\protect\citeauthoryear{{B{\"o}hringer} et~al.,}{{B{\"o}hringer} et~al.}{2004}]{Bohringer2004}
{B{\"o}hringer} H.,  et~al., 2004, \mn@doi [\aap] {10.1051/0004-6361:20034484}, \href {https://ui.adsabs.harvard.edu/abs/2004A&A...425..367B} {425, 367}

\bibitem[\protect\citeauthoryear{{Booth} \& {Schaye}}{{Booth} \& {Schaye}}{2009}]{BoothSchaye2009}
{Booth} C.~M.,  {Schaye} J.,  2009, \mn@doi [\mnras] {10.1111/j.1365-2966.2009.15043.x}, \href {https://ui.adsabs.harvard.edu/abs/2009MNRAS.398...53B} {398, 53}

\bibitem[\protect\citeauthoryear{{Borrow}, {Schaller}, {Bower}  \& {Schaye}}{{Borrow} et~al.}{2022}]{Borrow2022}
{Borrow} J.,  {Schaller} M.,  {Bower} R.~G.,   {Schaye} J.,  2022, \mn@doi [\mnras] {10.1093/mnras/stab3166}, \href {https://ui.adsabs.harvard.edu/abs/2022MNRAS.511.2367B} {511, 2367}

\bibitem[\protect\citeauthoryear{{Bulbul} et~al.,}{{Bulbul} et~al.}{2019}]{Bulbul2019}
{Bulbul} E.,  et~al., 2019, \mn@doi [\apj] {10.3847/1538-4357/aaf230}, \href {https://ui.adsabs.harvard.edu/abs/2019ApJ...871...50B} {871, 50}

\bibitem[\protect\citeauthoryear{{Chaikin}, {Schaye}, {Schaller}, {Bah{\'e}}, {Nobels}  \& {Ploeckinger}}{{Chaikin} et~al.}{2022}]{Chaikin2022}
{Chaikin} E.,  {Schaye} J.,  {Schaller} M.,  {Bah{\'e}} Y.~M.,  {Nobels} F. S.~J.,   {Ploeckinger} S.,  2022, \mn@doi [\mnras] {10.1093/mnras/stac1132}, \href {https://ui.adsabs.harvard.edu/abs/2022MNRAS.514..249C} {514, 249}

\bibitem[\protect\citeauthoryear{{Croston} et~al.,}{{Croston} et~al.}{2008}]{Croston2008}
{Croston} J.~H.,  et~al., 2008, \mn@doi [\aap] {10.1051/0004-6361:20079154}, \href {https://ui.adsabs.harvard.edu/abs/2008A&A...487..431C} {487, 431}

\bibitem[\protect\citeauthoryear{{De Grandi} \& {Molendi}}{{De Grandi} \& {Molendi}}{2001}]{DeGrandi2001}
{De Grandi} S.,  {Molendi} S.,  2001, \mn@doi [\apj] {10.1086/320098}, \href {https://ui.adsabs.harvard.edu/abs/2001ApJ...551..153D} {551, 153}

\bibitem[\protect\citeauthoryear{{Eckert}, {Molendi}  \& {Paltani}}{{Eckert} et~al.}{2011}]{Eckert2011}
{Eckert} D.,  {Molendi} S.,   {Paltani} S.,  2011, \mn@doi [\aap] {10.1051/0004-6361/201015856}, \href {https://ui.adsabs.harvard.edu/abs/2011A&A...526A..79E} {526, A79}

\bibitem[\protect\citeauthoryear{{Eckert} et~al.,}{{Eckert} et~al.}{2016}]{Eckert2016}
{Eckert} D.,  et~al., 2016, \mn@doi [\aap] {10.1051/0004-6361/201527293}, \href {https://ui.adsabs.harvard.edu/abs/2016A&A...592A..12E} {592, A12}

\bibitem[\protect\citeauthoryear{{Elahi}, {Ca{\~n}as}, {Poulton}, {Tobar}, {Willis}, {Lagos}, {Power}  \& {Robotham}}{{Elahi} et~al.}{2019}]{Elahi2019vr}
{Elahi} P.~J.,  {Ca{\~n}as} R.,  {Poulton} R. J.~J.,  {Tobar} R.~J.,  {Willis} J.~S.,  {Lagos} C. d.~P.,  {Power} C.,   {Robotham} A. S.~G.,  2019, \mn@doi [\pasa] {10.1017/pasa.2019.12}, \href {https://ui.adsabs.harvard.edu/abs/2019PASA...36...21E} {36, e021}

\bibitem[\protect\citeauthoryear{{Elbers}, {Frenk}, {Jenkins}, {Li}  \& {Pascoli}}{{Elbers} et~al.}{2021}]{Elbers2021}
{Elbers} W.,  {Frenk} C.~S.,  {Jenkins} A.,  {Li} B.,   {Pascoli} S.,  2021, \mn@doi [\mnras] {10.1093/mnras/stab2260}, \href {https://ui.adsabs.harvard.edu/abs/2021MNRAS.507.2614E} {507, 2614}

\bibitem[\protect\citeauthoryear{{Elbers}, {Frenk}, {Jenkins}, {Li}  \& {Pascoli}}{{Elbers} et~al.}{2022}]{Elbers2022}
{Elbers} W.,  {Frenk} C.~S.,  {Jenkins} A.,  {Li} B.,   {Pascoli} S.,  2022, \mn@doi [\mnras] {10.1093/mnras/stac2365}, \href {https://ui.adsabs.harvard.edu/abs/2022MNRAS.516.3821E} {516, 3821}

\bibitem[\protect\citeauthoryear{{Ettori}, {Donnarumma}, {Pointecouteau}, {Reiprich}, {Giodini}, {Lovisari}  \& {Schmidt}}{{Ettori} et~al.}{2013}]{Ettori2013}
{Ettori} S.,  {Donnarumma} A.,  {Pointecouteau} E.,  {Reiprich} T.~H.,  {Giodini} S.,  {Lovisari} L.,   {Schmidt} R.~W.,  2013, \mn@doi [\ssr] {10.1007/s11214-013-9976-7}, \href {https://ui.adsabs.harvard.edu/abs/2013SSRv..177..119E} {177, 119}

\bibitem[\protect\citeauthoryear{{Faucher-Gigu{\`e}re}}{{Faucher-Gigu{\`e}re}}{2020}]{FG20}
{Faucher-Gigu{\`e}re} C.-A.,  2020, \mn@doi [\mnras] {10.1093/mnras/staa302}, \href {https://ui.adsabs.harvard.edu/abs/2020MNRAS.493.1614F} {493, 1614}

\bibitem[\protect\citeauthoryear{{Ferland} et~al.,}{{Ferland} et~al.}{2017}]{Ferland2017}
{Ferland} G.~J.,  et~al., 2017, \mn@doi [\rmxaa] {10.48550/arXiv.1705.10877}, \href {https://ui.adsabs.harvard.edu/abs/2017RMxAA..53..385F} {53, 385}

\bibitem[\protect\citeauthoryear{{Gaspari} et~al.,}{{Gaspari} et~al.}{2019}]{Gaspari2019}
{Gaspari} M.,  et~al., 2019, \mn@doi [\apj] {10.3847/1538-4357/ab3c5d}, \href {https://ui.adsabs.harvard.edu/abs/2019ApJ...884..169G} {884, 169}

\bibitem[\protect\citeauthoryear{{Ghirardini} et~al.,}{{Ghirardini} et~al.}{2019}]{Ghirardini2019}
{Ghirardini} V.,  et~al., 2019, \mn@doi [\aap] {10.1051/0004-6361/201833325}, \href {https://ui.adsabs.harvard.edu/abs/2019A&A...621A..41G} {621, A41}

\bibitem[\protect\citeauthoryear{{Ghirardini} et~al.,}{{Ghirardini} et~al.}{2021}]{Ghirardini2021}
{Ghirardini} V.,  et~al., 2021, \mn@doi [\apj] {10.3847/1538-4357/abc68d}, \href {https://ui.adsabs.harvard.edu/abs/2021ApJ...910...14G} {910, 14}

\bibitem[\protect\citeauthoryear{{Ghizzardi} et~al.,}{{Ghizzardi} et~al.}{2021}]{Ghizzardi2021}
{Ghizzardi} S.,  et~al., 2021, \mn@doi [\aap] {10.1051/0004-6361/202038501}, \href {https://ui.adsabs.harvard.edu/abs/2021A&A...646A..92G} {646, A92}

\bibitem[\protect\citeauthoryear{{Gianfagna} et~al.,}{{Gianfagna} et~al.}{2021}]{Gianfagna2021}
{Gianfagna} G.,  et~al., 2021, \mn@doi [\mnras] {10.1093/mnras/stab308}, \href {https://ui.adsabs.harvard.edu/abs/2021MNRAS.502.5115G} {502, 5115}

\bibitem[\protect\citeauthoryear{{Hahn}, {Rampf}  \& {Uhlemann}}{{Hahn} et~al.}{2021}]{Hahn2021}
{Hahn} O.,  {Rampf} C.,   {Uhlemann} C.,  2021, \mn@doi [\mnras] {10.1093/mnras/staa3773}, \href {https://ui.adsabs.harvard.edu/abs/2021MNRAS.503..426H} {503, 426}

\bibitem[\protect\citeauthoryear{{Henson}, {Barnes}, {Kay}, {McCarthy}  \& {Schaye}}{{Henson} et~al.}{2017}]{Henson2017}
{Henson} M.~A.,  {Barnes} D.~J.,  {Kay} S.~T.,  {McCarthy} I.~G.,   {Schaye} J.,  2017, \mn@doi [\mnras] {10.1093/mnras/stw2899}, \href {https://ui.adsabs.harvard.edu/abs/2017MNRAS.465.3361H} {465, 3361}

\bibitem[\protect\citeauthoryear{{Hilton} et~al.,}{{Hilton} et~al.}{2021}]{Hilton2021}
{Hilton} M.,  et~al., 2021, \mn@doi [\apjs] {10.3847/1538-4365/abd023}, \href {https://ui.adsabs.harvard.edu/abs/2021ApJS..253....3H} {253, 3}

\bibitem[\protect\citeauthoryear{{Hoekstra}, {Herbonnet}, {Muzzin}, {Babul}, {Mahdavi}, {Viola}  \& {Cacciato}}{{Hoekstra} et~al.}{2015}]{Hoekstra2015}
{Hoekstra} H.,  {Herbonnet} R.,  {Muzzin} A.,  {Babul} A.,  {Mahdavi} A.,  {Viola} M.,   {Cacciato} M.,  2015, \mn@doi [\mnras] {10.1093/mnras/stv275}, \href {https://ui.adsabs.harvard.edu/abs/2015MNRAS.449..685H} {449, 685}

\bibitem[\protect\citeauthoryear{{Hudson}, {Mittal}, {Reiprich}, {Nulsen}, {Andernach}  \& {Sarazin}}{{Hudson} et~al.}{2010}]{hudson2010}
{Hudson} D.~S.,  {Mittal} R.,  {Reiprich} T.~H.,  {Nulsen} P.~E.~J.,  {Andernach} H.,   {Sarazin} C.~L.,  2010, \mn@doi [\aap] {10.1051/0004-6361/200912377}, \href {https://ui.adsabs.harvard.edu/abs/2010A&A...513A..37H} {513, A37}

\bibitem[\protect\citeauthoryear{{Hu{\v{s}}ko}, {Lacey}, {Schaye}, {Schaller}  \& {Nobels}}{{Hu{\v{s}}ko} et~al.}{2022}]{Husko2022}
{Hu{\v{s}}ko} F.,  {Lacey} C.~G.,  {Schaye} J.,  {Schaller} M.,   {Nobels} F. S.~J.,  2022, \mn@doi [\mnras] {10.1093/mnras/stac2278}, \href {https://ui.adsabs.harvard.edu/abs/2022MNRAS.516.3750H} {516, 3750}

\bibitem[\protect\citeauthoryear{{Jennings} \& {Dav{\'e}}}{{Jennings} \& {Dav{\'e}}}{2023}]{Jennings2023}
{Jennings} F.,  {Dav{\'e}} R.,  2023, \mn@doi [\mnras] {10.1093/mnras/stad2666}, \href {https://ui.adsabs.harvard.edu/abs/2023MNRAS.526.1367J} {526, 1367}

\bibitem[\protect\citeauthoryear{{Kaiser}}{{Kaiser}}{1986}]{Kaiser1986}
{Kaiser} N.,  1986, \mn@doi [\mnras] {10.1093/mnras/222.2.323}, \href {https://ui.adsabs.harvard.edu/abs/1986MNRAS.222..323K} {222, 323}

\bibitem[\protect\citeauthoryear{{Kaiser}}{{Kaiser}}{1991}]{Kaiser1991}
{Kaiser} N.,  1991, \mn@doi [\apj] {10.1086/170768}, \href {https://ui.adsabs.harvard.edu/abs/1991ApJ...383..104K} {383, 104}

\bibitem[\protect\citeauthoryear{{Kugel} et~al.,}{{Kugel} et~al.}{2023}]{Kugel2023}
{Kugel} R.,  et~al., 2023, \mn@doi [\mnras] {10.1093/mnras/stad2540}, \href {https://ui.adsabs.harvard.edu/abs/2023MNRAS.526.6103K} {526, 6103}

\bibitem[\protect\citeauthoryear{{Lakhchaura}, {Saini}  \& {Sharma}}{{Lakhchaura} et~al.}{2016}]{Lakhchaura2016}
{Lakhchaura} K.,  {Saini} T.~D.,   {Sharma} P.,  2016, \mn@doi [\mnras] {10.1093/mnras/stw1062}, \href {https://ui.adsabs.harvard.edu/abs/2016MNRAS.460.2625L} {460, 2625}

\bibitem[\protect\citeauthoryear{{Le Brun}, {McCarthy}, {Schaye}  \& {Ponman}}{{Le Brun} et~al.}{2014}]{Lebrun2014}
{Le Brun} A. M.~C.,  {McCarthy} I.~G.,  {Schaye} J.,   {Ponman} T.~J.,  2014, \mn@doi [\mnras] {10.1093/mnras/stu608}, \href {https://ui.adsabs.harvard.edu/abs/2014MNRAS.441.1270L} {441, 1270}

\bibitem[\protect\citeauthoryear{{Lehle}, {Nelson}, {Pillepich}, {Truong}  \& {Rohr}}{{Lehle} et~al.}{2023}]{Lehle2023}
{Lehle} K.,  {Nelson} D.,  {Pillepich} A.,  {Truong} N.,   {Rohr} E.,  2023, \mn@doi [arXiv e-prints] {10.48550/arXiv.2311.06333}, \href {https://ui.adsabs.harvard.edu/abs/2023arXiv231106333L} {p. arXiv:2311.06333}

\bibitem[\protect\citeauthoryear{{Li} et~al.,}{{Li} et~al.}{2023}]{Li2023}
{Li} Q.,  et~al., 2023, \mn@doi [\mnras] {10.1093/mnras/stad1521}, \href {https://ui.adsabs.harvard.edu/abs/2023MNRAS.523.1228L} {523, 1228}

\bibitem[\protect\citeauthoryear{{Lin}, {McDonald}, {Benson}  \& {Miller}}{{Lin} et~al.}{2015}]{Lin2015}
{Lin} H.~W.,  {McDonald} M.,  {Benson} B.,   {Miller} E.,  2015, \mn@doi [\apj] {10.1088/0004-637X/802/1/34}, \href {https://ui.adsabs.harvard.edu/abs/2015ApJ...802...34L} {802, 34}

\bibitem[\protect\citeauthoryear{{Liu} et~al.,}{{Liu} et~al.}{2022}]{Liu2022}
{Liu} A.,  et~al., 2022, \mn@doi [\aap] {10.1051/0004-6361/202141120}, \href {https://ui.adsabs.harvard.edu/abs/2022A&A...661A...2L} {661, A2}

\bibitem[\protect\citeauthoryear{{Lovisari} \& {Reiprich}}{{Lovisari} \& {Reiprich}}{2019}]{Lovisari2019}
{Lovisari} L.,  {Reiprich} T.~H.,  2019, \mn@doi [\mnras] {10.1093/mnras/sty3130}, \href {https://ui.adsabs.harvard.edu/abs/2019MNRAS.483..540L} {483, 540}

\bibitem[\protect\citeauthoryear{{Lovisari}, {Reiprich}  \& {Schellenberger}}{{Lovisari} et~al.}{2015}]{Lovisari2015}
{Lovisari} L.,  {Reiprich} T.~H.,   {Schellenberger} G.,  2015, \mn@doi [\aap] {10.1051/0004-6361/201423954}, \href {https://ui.adsabs.harvard.edu/abs/2015A&A...573A.118L} {573, A118}

\bibitem[\protect\citeauthoryear{{Lovisari} et~al.,}{{Lovisari} et~al.}{2017}]{Lovisari2017}
{Lovisari} L.,  et~al., 2017, \mn@doi [\apj] {10.3847/1538-4357/aa855f}, \href {https://ui.adsabs.harvard.edu/abs/2017ApJ...846...51L} {846, 51}

\bibitem[\protect\citeauthoryear{{Lovisari} et~al.,}{{Lovisari} et~al.}{2020}]{Lovisari2020}
{Lovisari} L.,  et~al., 2020, \mn@doi [\apj] {10.3847/1538-4357/ab7997}, \href {https://ui.adsabs.harvard.edu/abs/2020ApJ...892..102L} {892, 102}

\bibitem[\protect\citeauthoryear{{Lovisari}, {Ettori}, {Gaspari}  \& {Giles}}{{Lovisari} et~al.}{2021}]{Lovisari2021}
{Lovisari} L.,  {Ettori} S.,  {Gaspari} M.,   {Giles} P.~A.,  2021, \mn@doi [Universe] {10.3390/universe7050139}, \href {https://ui.adsabs.harvard.edu/abs/2021Univ....7..139L} {7, 139}

\bibitem[\protect\citeauthoryear{{Mahdavi}, {Hoekstra}, {Babul}, {Bildfell}, {Jeltema}  \& {Henry}}{{Mahdavi} et~al.}{2013}]{Mahdavi2013}
{Mahdavi} A.,  {Hoekstra} H.,  {Babul} A.,  {Bildfell} C.,  {Jeltema} T.,   {Henry} J.~P.,  2013, \mn@doi [\apj] {10.1088/0004-637X/767/2/116}, \href {https://ui.adsabs.harvard.edu/abs/2013ApJ...767..116M} {767, 116}

\bibitem[\protect\citeauthoryear{{Marigo}}{{Marigo}}{2001}]{Marigo2001}
{Marigo} P.,  2001, \mn@doi [\aap] {10.1051/0004-6361:20000247}, \href {https://ui.adsabs.harvard.edu/abs/2001A&A...370..194M} {370, 194}

\bibitem[\protect\citeauthoryear{{Maughan}, {Giles}, {Randall}, {Jones}  \& {Forman}}{{Maughan} et~al.}{2012}]{Maughan2012}
{Maughan} B.~J.,  {Giles} P.~A.,  {Randall} S.~W.,  {Jones} C.,   {Forman} W.~R.,  2012, \mn@doi [\mnras] {10.1111/j.1365-2966.2012.20419.x}, \href {https://ui.adsabs.harvard.edu/abs/2012MNRAS.421.1583M} {421, 1583}

\bibitem[\protect\citeauthoryear{{Mazzotta}, {Rasia}, {Moscardini}  \& {Tormen}}{{Mazzotta} et~al.}{2004}]{Mazzotta2004}
{Mazzotta} P.,  {Rasia} E.,  {Moscardini} L.,   {Tormen} G.,  2004, \mn@doi [\mnras] {10.1111/j.1365-2966.2004.08167.x}, \href {https://ui.adsabs.harvard.edu/abs/2004MNRAS.354...10M} {354, 10}

\bibitem[\protect\citeauthoryear{{McCarthy}, {Schaye}, {Bower}, {Ponman}, {Booth}, {Dalla Vecchia}  \& {Springel}}{{McCarthy} et~al.}{2011}]{Mccarthy2011}
{McCarthy} I.~G.,  {Schaye} J.,  {Bower} R.~G.,  {Ponman} T.~J.,  {Booth} C.~M.,  {Dalla Vecchia} C.,   {Springel} V.,  2011, \mn@doi [\mnras] {10.1111/j.1365-2966.2010.18033.x}, \href {https://ui.adsabs.harvard.edu/abs/2011MNRAS.412.1965M} {412, 1965}

\bibitem[\protect\citeauthoryear{{McCarthy}, {Schaye}, {Bird}  \& {Le Brun}}{{McCarthy} et~al.}{2017}]{Mccarthy2017}
{McCarthy} I.~G.,  {Schaye} J.,  {Bird} S.,   {Le Brun} A. M.~C.,  2017, \mn@doi [\mnras] {10.1093/mnras/stw2792}, \href {https://ui.adsabs.harvard.edu/abs/2017MNRAS.465.2936M} {465, 2936}

\bibitem[\protect\citeauthoryear{{McDonald} et~al.,}{{McDonald} et~al.}{2013}]{McDonald2013}
{McDonald} M.,  et~al., 2013, \mn@doi [\apj] {10.1088/0004-637X/774/1/23}, \href {https://ui.adsabs.harvard.edu/abs/2013ApJ...774...23M} {774, 23}

\bibitem[\protect\citeauthoryear{{McDonald} et~al.,}{{McDonald} et~al.}{2014}]{McDonald2014}
{McDonald} M.,  et~al., 2014, \mn@doi [\apj] {10.1088/0004-637X/794/1/67}, \href {https://ui.adsabs.harvard.edu/abs/2014ApJ...794...67M} {794, 67}

\bibitem[\protect\citeauthoryear{{McDonald} et~al.,}{{McDonald} et~al.}{2017}]{McDonald2017}
{McDonald} M.,  et~al., 2017, \mn@doi [\apj] {10.3847/1538-4357/aa7740}, \href {https://ui.adsabs.harvard.edu/abs/2017ApJ...843...28M} {843, 28}

\bibitem[\protect\citeauthoryear{{Migkas}, {Schellenberger}, {Reiprich}, {Pacaud}, {Ramos-Ceja}  \& {Lovisari}}{{Migkas} et~al.}{2020}]{Migkas2020}
{Migkas} K.,  {Schellenberger} G.,  {Reiprich} T.~H.,  {Pacaud} F.,  {Ramos-Ceja} M.~E.,   {Lovisari} L.,  2020, \mn@doi [\aap] {10.1051/0004-6361/201936602}, \href {https://ui.adsabs.harvard.edu/abs/2020A&A...636A..15M} {636, A15}

\bibitem[\protect\citeauthoryear{{Mu{\~n}oz-Echeverr{\'\i}a} et~al.,}{{Mu{\~n}oz-Echeverr{\'\i}a} et~al.}{2023}]{Munoz2023}
{Mu{\~n}oz-Echeverr{\'\i}a} M.,  et~al., 2023, \mn@doi [arXiv e-prints] {10.48550/arXiv.2312.01154}, \href {https://ui.adsabs.harvard.edu/abs/2023arXiv231201154M} {p. arXiv:2312.01154}

\bibitem[\protect\citeauthoryear{{Mukai}}{{Mukai}}{1993}]{Mukai1993}
{Mukai} K.,  1993, Legacy, \href {https://ui.adsabs.harvard.edu/abs/1993Legac...3...21M} {3, 21}

\bibitem[\protect\citeauthoryear{{Nagai}, {Vikhlinin}  \& {Kravtsov}}{{Nagai} et~al.}{2007}]{Nagai2007}
{Nagai} D.,  {Vikhlinin} A.,   {Kravtsov} A.~V.,  2007, \mn@doi [\apj] {10.1086/509868}, \href {https://ui.adsabs.harvard.edu/abs/2007ApJ...655...98N} {655, 98}

\bibitem[\protect\citeauthoryear{{Nelson}, {Pillepich}, {Ayromlou}, {Lee}, {Lehle}, {Rohr}  \& {Truong}}{{Nelson} et~al.}{2023}]{Nelson2023}
{Nelson} D.,  {Pillepich} A.,  {Ayromlou} M.,  {Lee} W.,  {Lehle} K.,  {Rohr} E.,   {Truong} N.,  2023, arXiv e-prints, \href {https://ui.adsabs.harvard.edu/abs/2023arXiv231106338N} {p. arXiv:2311.06338}

\bibitem[\protect\citeauthoryear{{Neumann} \& {Arnaud}}{{Neumann} \& {Arnaud}}{1999}]{Neumann1999}
{Neumann} D.~M.,  {Arnaud} M.,  1999, \mn@doi [\aap] {10.48550/arXiv.astro-ph/9901092}, \href {https://ui.adsabs.harvard.edu/abs/1999A&A...348..711N} {348, 711}

\bibitem[\protect\citeauthoryear{{Pakmor} et~al.,}{{Pakmor} et~al.}{2023}]{Pakmor2023}
{Pakmor} R.,  et~al., 2023, \mn@doi [\mnras] {10.1093/mnras/stac3620}, \href {https://ui.adsabs.harvard.edu/abs/2023MNRAS.524.2539P} {524, 2539}

\bibitem[\protect\citeauthoryear{{Pearce}, {Kay}, {Barnes}, {Bah{\'e}}  \& {Bower}}{{Pearce} et~al.}{2021}]{Pearce2021}
{Pearce} F.~A.,  {Kay} S.~T.,  {Barnes} D.~J.,  {Bah{\'e}} Y.~M.,   {Bower} R.~G.,  2021, \mn@doi [\mnras] {10.1093/mnras/stab2194}, \href {https://ui.adsabs.harvard.edu/abs/2021MNRAS.507.1606P} {507, 1606}

\bibitem[\protect\citeauthoryear{{Pearson} et~al.,}{{Pearson} et~al.}{2017}]{Pearson2017}
{Pearson} R.~J.,  et~al., 2017, \mn@doi [\mnras] {10.1093/mnras/stx1081}, \href {https://ui.adsabs.harvard.edu/abs/2017MNRAS.469.3489P} {469, 3489}

\bibitem[\protect\citeauthoryear{{Pereira} et~al.,}{{Pereira} et~al.}{2018}]{Pereira2018}
{Pereira} M. E.~S.,  et~al., 2018, \mn@doi [\mnras] {10.1093/mnras/stx2831}, \href {https://ui.adsabs.harvard.edu/abs/2018MNRAS.474.1361P} {474, 1361}

\bibitem[\protect\citeauthoryear{{Planck Collaboration} et~al.,}{{Planck Collaboration} et~al.}{2016}]{Planck2016}
{Planck Collaboration} et~al., 2016, \mn@doi [\aap] {10.1051/0004-6361/201525833}, \href {https://ui.adsabs.harvard.edu/abs/2016A&A...594A..24P} {594, A24}

\bibitem[\protect\citeauthoryear{{Ploeckinger} \& {Schaye}}{{Ploeckinger} \& {Schaye}}{2020}]{ploeckinger2020}
{Ploeckinger} S.,  {Schaye} J.,  2020, \mn@doi [\mnras] {10.1093/mnras/staa2172}, \href {https://ui.adsabs.harvard.edu/abs/2020MNRAS.497.4857P} {497, 4857}

\bibitem[\protect\citeauthoryear{{Ponman}, {Sanderson}  \& {Finoguenov}}{{Ponman} et~al.}{2003}]{Ponman2003}
{Ponman} T.~J.,  {Sanderson} A.~J.~R.,   {Finoguenov} A.,  2003, \mn@doi [\mnras] {10.1046/j.1365-8711.2003.06677.x}, \href {https://ui.adsabs.harvard.edu/abs/2003MNRAS.343..331P} {343, 331}

\bibitem[\protect\citeauthoryear{{Portinari}, {Chiosi}  \& {Bressan}}{{Portinari} et~al.}{1998}]{Portinari1998}
{Portinari} L.,  {Chiosi} C.,   {Bressan} A.,  1998, \mn@doi [\aap] {10.48550/arXiv.astro-ph/9711337}, \href {https://ui.adsabs.harvard.edu/abs/1998A&A...334..505P} {334, 505}

\bibitem[\protect\citeauthoryear{{Pratt}, {Croston}, {Arnaud}  \& {B{\"o}hringer}}{{Pratt} et~al.}{2009}]{Pratt2009}
{Pratt} G.~W.,  {Croston} J.~H.,  {Arnaud} M.,   {B{\"o}hringer} H.,  2009, \mn@doi [\aap] {10.1051/0004-6361/200810994}, \href {https://ui.adsabs.harvard.edu/abs/2009A&A...498..361P} {498, 361}

\bibitem[\protect\citeauthoryear{{Rasia} et~al.,}{{Rasia} et~al.}{2012}]{Rasia2012}
{Rasia} E.,  et~al., 2012, \mn@doi [New Journal of Physics] {10.1088/1367-2630/14/5/055018}, \href {https://ui.adsabs.harvard.edu/abs/2012NJPh...14e5018R} {14, 055018}

\bibitem[\protect\citeauthoryear{{Robson} \& {Dav{\'e}}}{{Robson} \& {Dav{\'e}}}{2023}]{Robson2023}
{Robson} D.,  {Dav{\'e}} R.,  2023, \mn@doi [\mnras] {10.1093/mnras/stac2982}, \href {https://ui.adsabs.harvard.edu/abs/2023MNRAS.518.5826R} {518, 5826}

\bibitem[\protect\citeauthoryear{{Rossetti}, {Gastaldello}, {Eckert}, {Della Torre}, {Pantiri}, {Cazzoletti}  \& {Molendi}}{{Rossetti} et~al.}{2017}]{Rossetti2017}
{Rossetti} M.,  {Gastaldello} F.,  {Eckert} D.,  {Della Torre} M.,  {Pantiri} G.,  {Cazzoletti} P.,   {Molendi} S.,  2017, \mn@doi [\mnras] {10.1093/mnras/stx493}, \href {https://ui.adsabs.harvard.edu/abs/2017MNRAS.468.1917R} {468, 1917}

\bibitem[\protect\citeauthoryear{{Ruppin}, {McDonald}, {Bleem}, {Allen}, {Benson}, {Calzadilla}, {Khullar}  \& {Floyd}}{{Ruppin} et~al.}{2021}]{Ruppin2021}
{Ruppin} F.,  {McDonald} M.,  {Bleem} L.~E.,  {Allen} S.~W.,  {Benson} B.~A.,  {Calzadilla} M.,  {Khullar} G.,   {Floyd} B.,  2021, \mn@doi [\apj] {10.3847/1538-4357/ac0bba}, \href {https://ui.adsabs.harvard.edu/abs/2021ApJ...918...43R} {918, 43}

\bibitem[\protect\citeauthoryear{{Rykoff} et~al.,}{{Rykoff} et~al.}{2014}]{Rykoff2014}
{Rykoff} E.~S.,  et~al., 2014, \mn@doi [\apj] {10.1088/0004-637X/785/2/104}, \href {https://ui.adsabs.harvard.edu/abs/2014ApJ...785..104R} {785, 104}

\bibitem[\protect\citeauthoryear{{Sayers} et~al.,}{{Sayers} et~al.}{2013}]{Sayers2013}
{Sayers} J.,  et~al., 2013, \mn@doi [\apj] {10.1088/0004-637X/768/2/177}, \href {https://ui.adsabs.harvard.edu/abs/2013ApJ...768..177S} {768, 177}

\bibitem[\protect\citeauthoryear{{Schaller} et~al.,}{{Schaller} et~al.}{2023}]{Schaller2023}
{Schaller} M.,  et~al., 2023, \mn@doi [arXiv e-prints] {10.48550/arXiv.2305.13380}, \href {https://ui.adsabs.harvard.edu/abs/2023arXiv230513380S} {p. arXiv:2305.13380}

\bibitem[\protect\citeauthoryear{{Schaye} \& {Dalla Vecchia}}{{Schaye} \& {Dalla Vecchia}}{2008}]{Schaye2008}
{Schaye} J.,  {Dalla Vecchia} C.,  2008, \mn@doi [\mnras] {10.1111/j.1365-2966.2007.12639.x}, \href {https://ui.adsabs.harvard.edu/abs/2008MNRAS.383.1210S} {383, 1210}

\bibitem[\protect\citeauthoryear{{Schaye} et~al.,}{{Schaye} et~al.}{2010}]{Schaye2010}
{Schaye} J.,  et~al., 2010, \mn@doi [\mnras] {10.1111/j.1365-2966.2009.16029.x}, \href {https://ui.adsabs.harvard.edu/abs/2010MNRAS.402.1536S} {402, 1536}

\bibitem[\protect\citeauthoryear{Schaye et~al.,}{Schaye et~al.}{2015}]{Schaye2015}
Schaye J.,  et~al., 2015, \mn@doi [Monthly Notices of the Royal Astronomical Society] {10.1093/mnras/stu2058}, 446, 521

\bibitem[\protect\citeauthoryear{{Schaye} et~al.,}{{Schaye} et~al.}{2023}]{schaye2023}
{Schaye} J.,  et~al., 2023, \mn@doi [\mnras] {10.1093/mnras/stad2419}, \href {https://ui.adsabs.harvard.edu/abs/2023MNRAS.526.4978S} {526, 4978}

\bibitem[\protect\citeauthoryear{{Springel}, {Di Matteo}  \& {Hernquist}}{{Springel} et~al.}{2005}]{Springel2005}
{Springel} V.,  {Di Matteo} T.,   {Hernquist} L.,  2005, \mn@doi [\mnras] {10.1111/j.1365-2966.2005.09238.x}, \href {https://ui.adsabs.harvard.edu/abs/2005MNRAS.361..776S} {361, 776}

\bibitem[\protect\citeauthoryear{{Sun}, {Voit}, {Donahue}, {Jones}, {Forman}  \& {Vikhlinin}}{{Sun} et~al.}{2009}]{Sun2009}
{Sun} M.,  {Voit} G.~M.,  {Donahue} M.,  {Jones} C.,  {Forman} W.,   {Vikhlinin} A.,  2009, \mn@doi [\apj] {10.1088/0004-637X/693/2/1142}, \href {https://ui.adsabs.harvard.edu/abs/2009ApJ...693.1142S} {693, 1142}

\bibitem[\protect\citeauthoryear{{Sun}, {Sehgal}, {Voit}, {Donahue}, {Jones}, {Forman}, {Vikhlinin}  \& {Sarazin}}{{Sun} et~al.}{2011}]{Sun2011}
{Sun} M.,  {Sehgal} N.,  {Voit} G.~M.,  {Donahue} M.,  {Jones} C.,  {Forman} W.,  {Vikhlinin} A.,   {Sarazin} C.,  2011, \mn@doi [\apjl] {10.1088/2041-8205/727/2/L49}, \href {https://ui.adsabs.harvard.edu/abs/2011ApJ...727L..49S} {727, L49}

\bibitem[\protect\citeauthoryear{{Takey}, {Schwope}  \& {Lamer}}{{Takey} et~al.}{2013}]{Takey2013}
{Takey} A.,  {Schwope} A.,   {Lamer} G.,  2013, \mn@doi [\aap] {10.1051/0004-6361/201220213}, \href {https://ui.adsabs.harvard.edu/abs/2013A&A...558A..75T} {558, A75}

\bibitem[\protect\citeauthoryear{{Thielemann} et~al.,}{{Thielemann} et~al.}{2003}]{Thielemann2003}
{Thielemann} F.~K.,  et~al., 2003, in {Hillebrandt} W.,  {Leibundgut} B.,  eds, From Twilight to Highlight: The Physics of Supernovae. p.~331, \mn@doi{10.1007/10828549\_46}

\bibitem[\protect\citeauthoryear{{Towler}, {Kay}  \& {Altamura}}{{Towler} et~al.}{2023}]{Towler2023}
{Towler} I.,  {Kay} S.~T.,   {Altamura} E.,  2023, \mn@doi [\mnras] {10.1093/mnras/stad453}, \href {https://ui.adsabs.harvard.edu/abs/2023MNRAS.520.5845T} {520, 5845}

\bibitem[\protect\citeauthoryear{{Vikhlinin}}{{Vikhlinin}}{2006}]{Vikhlinin2006}
{Vikhlinin} A.,  2006, \mn@doi [\apj] {10.1086/500121}, \href {https://ui.adsabs.harvard.edu/abs/2006ApJ...640..710V} {640, 710}

\bibitem[\protect\citeauthoryear{{Vogelsberger} et~al.,}{{Vogelsberger} et~al.}{2018}]{Vogelsberger2018}
{Vogelsberger} M.,  et~al., 2018, \mn@doi [\mnras] {10.1093/mnras/stx2955}, \href {https://ui.adsabs.harvard.edu/abs/2018MNRAS.474.2073V} {474, 2073}

\bibitem[\protect\citeauthoryear{{Wiersma}, {Schaye}  \& {Smith}}{{Wiersma} et~al.}{2009a}]{wiersma2009}
{Wiersma} R. P.~C.,  {Schaye} J.,   {Smith} B.~D.,  2009a, \mn@doi [\mnras] {10.1111/j.1365-2966.2008.14191.x}, \href {https://ui.adsabs.harvard.edu/abs/2009MNRAS.393...99W} {393, 99}

\bibitem[\protect\citeauthoryear{{Wiersma}, {Schaye}, {Theuns}, {Dalla Vecchia}  \& {Tornatore}}{{Wiersma} et~al.}{2009b}]{Wiersma_stellarmassloss2009}
{Wiersma} R. P.~C.,  {Schaye} J.,  {Theuns} T.,  {Dalla Vecchia} C.,   {Tornatore} L.,  2009b, \mn@doi [\mnras] {10.1111/j.1365-2966.2009.15331.x}, \href {https://ui.adsabs.harvard.edu/abs/2009MNRAS.399..574W} {399, 574}

\bibitem[\protect\citeauthoryear{{von der Linden} et~al.,}{{von der Linden} et~al.}{2014}]{vonderlinden2014}
{von der Linden} A.,  et~al., 2014, \mn@doi [\mnras] {10.1093/mnras/stt1945}, \href {https://ui.adsabs.harvard.edu/abs/2014MNRAS.439....2V} {439, 2}

\makeatother
\end{thebibliography}

% Alternatively you could enter them by hand, like this:
% This method is tedious and prone to error if you have lots of references
%\begin{thebibliography}{99}
%\bibitem[\protect\citeauthoryear{Author}{2012}]{Author2012}
%Author A.~N., 2013, Journal of Improbable Astronomy, 1, 1
%\bibitem[\protect\citeauthoryear{Others}{2013}]{Others2013}
%Others S., 2012, Journal of Interesting Stuff, 17, 198
%\end{thebibliography}

%%%%%%%%%%%%%%%%%%%%%%%%%%%%%%%%%%%%%%%%%%%%%%%%%%

%%%%%%%%%%%%%%%%% APPENDICES %%%%%%%%%%%%%%%%%%%%%

\appendix

\section{The evolution of scaling relations} \label{sec:appendix_scaling_relations}

FLAMINGO has been calibrated to match observed gas fractions at $\mathrm{R_{\rm 500c}}$ for clusters at $z \approx 0.1 - 0.3$ and $M_\text{500c} = 10^{13.5} - 10^{14.36}~\text{M}_\odot$, and we have shown here and in \citet{schaye2023} that this leads to a reproduction of the observed $z=0$ scaling relations. Here we show that this holds even at much higher redshifts.

\begin{figure*}
	\includegraphics[width=\linewidth]{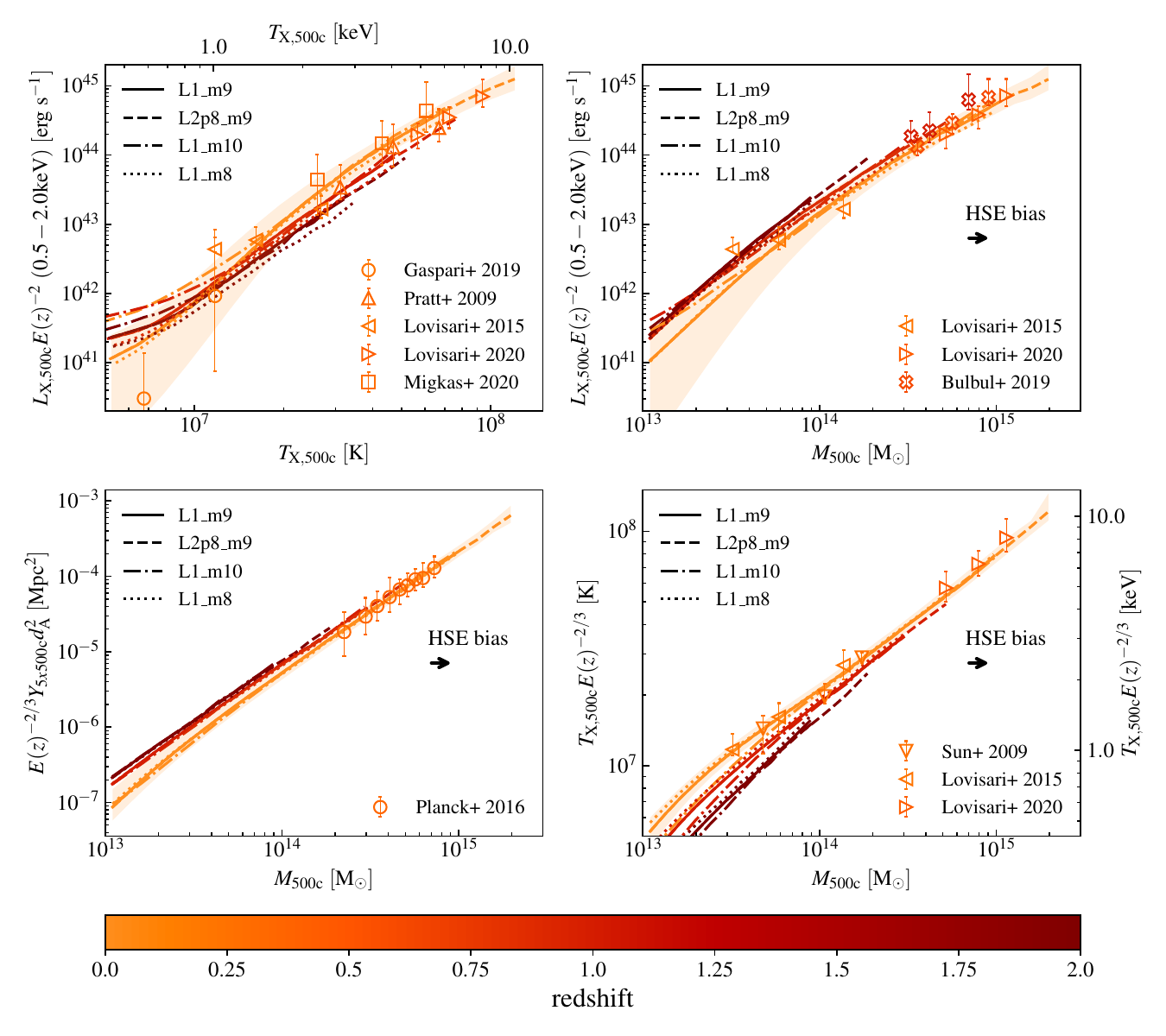}
    \caption{Evolution of cluster scaling relations for haloes with $M_{500\mathrm{c}} > 10^{13} \mathrm{M_\odot}$, for the fiducial model run at different numerical resolutions and box sizes (different lines). The lines are the median relations between X-ray luminosity and temperature (top left), X-ray luminosity and halo mass (top right), temperature and halo mass (bottom right) and thermal Sunyaev-Zel'dovich effect Compton Y and halo mass (bottom left). Luminosities, temperatures and masses are measured within $R_{\rm 500c}$ while Compton Y is measured within $5R_{\rm 500c}$. The solid, dashed, dot-dashed, and dotted lines show the results for different box sizes and resolutions. Comparisons are made with X-ray data from \citet[][$0.08 < z < 0.15$]{Pratt2009}, \citet[][$z < 0.035$]{Lovisari2015}, \citet[][$0.059<z<0.546$]{Lovisari2020}, \citet[][$0.2<z<1.5$]{Bulbul2019}, \citet[][$z < 0.04$]{Gaspari2019}, and \citet[][$z < 0.3$]{Migkas2020}, and with SZE data for $z<0.25$ clusters from \citet{Planck2016} compiled by \citet{Mccarthy2017}. We multiply the thermodynamic quantities by $E(z)^\alpha$, where $E(z)\equiv H(z)/H_0$, and the value of $\alpha$ is indicated in each panel, to correct the data to $z=0$ assuming self-similar evolution. The shaded region ($z=0$ only) and the observational error bars indicate the 16th and 84th percentiles of the sample. Observational data is colored according to the median redshift. The convergence with resolution and box size is excellent. There are only small deviations from self-similar evolution. }
    \label{fig:scaling_relations_evolution}
\end{figure*}

Figure \ref{fig:scaling_relations_evolution} shows the median X-ray luminosity-temperature (top left), luminosity-mass (top right), temperature-mass (bottom right), and Compton Y - mass\footnote{The Compton Y parameter within radius $R$ is defined as $\mathrm{Y}(<R) d_{\rm A}^2(z) = \frac{\sigma_{\rm T}k_{\rm B}}{m_{\rm e}c^2} \int_{< R} n_{\rm e}T \mathrm{d}V$} (bottom left) relations for all groups and clusters in the fiducial model for different box sizes and resolutions. Temperatures are mass-weighted and include all particles with $T>10^5 \mathrm{K}$, we have compared with emission-weighted temperatures and found the differences to be negligble for these global properties. The X-ray luminosity is scaled by $E(z)^{-2}$ ($E(z) = H(z) / H_0$), and the Compton-Y parameter and temperature have been scaled by $E(z)^{-2/3}$ to take out the expected self-similar redshift evolution. The small horizontal arrows indicate the systematic shift applied to data with hydrostatic-equilibrium inferred masses, which corresponds to the value found for the hydrostatic mass bias (0.743) during the calibration in \citet{Kugel2023}, which agrees with their assumed priors based on the observations of \citet{Hoekstra2015} and \citet{Eckert2016}. We neglect any corrections on the observed quantities due to the hydrostatic bias as these are negligible (e.g. the X-Ray luminosity will be dominated by radii much smaller than $R_{\rm 500c}$)

The observed X-ray luminosities have been shifted to the 0.5-2.0 keV band using PIMMS\footnote{\url{https://heasarc.gsfc.nasa.gov/docs/software/tools/pimms.html}} \citep{Mukai1993}. The SZE data has been compiled by \cite{Mccarthy2017} from the clusters observed by \citet{Planck2016}. Observational data was grouped into a limited number of bins per data set, where the error bars show the scatter between individual objects in that bin.

The different colors, showing redshifts $z = 0.0, 1.0, 2.0$, indicate that there is almost no evolution in the scaling relations after accounting for self-similar evolution \citep{Kaiser1986, Kaiser1991}, though the mass-temperature relation shows a slight trend towards lower temperatures at fixed mass when the redshift increases. This is similar to what was found by \citet{Barnes2017}, who ascribe it to larger non-thermal pressure from bulk gas motion at higher redshift. In agreement with previous observational studies \citep[e.g.][]{Maughan2012}, the evolution of the scaling relations is consistent with self-similar expectations even though the slopes of the relations are very different from self-similar expectations.

All four scaling relations show an excellent match to the observations, which are mostly at low redshift, across the entire mass range. Even the lowest resolution, simulation (L1\_m10) is nearly identical to the highest resolution simulation (L1\_m8) at all redshifts, with only a slight deviation visible at $M_{500\mathrm{c}} < 3 \times 10^{13} \mathrm{M_\odot}$ or, equivalently, $T_{500\mathrm{c}} < 2 \times 10^7 \mathrm{K}$. At fixed resolution (m9), there is near perfect agreement between different box sizes, and the larger volume smoothly extends the scaling relation to higher masses.

In summary, the simulated cluster scaling relations are numerically converged and agree well with observations. Their evolution is close to self-similar.

\section{Projection effects} \label{sec:appendix_projection}
When observing galaxy clusters, the emission projected along the line of sight is observed. In the analysis of observations, a (model-dependent) deprojection procedure is performed such that analytic 3D profiles can be fitted to the results. 

\begin{figure}
    \centering
    \includegraphics{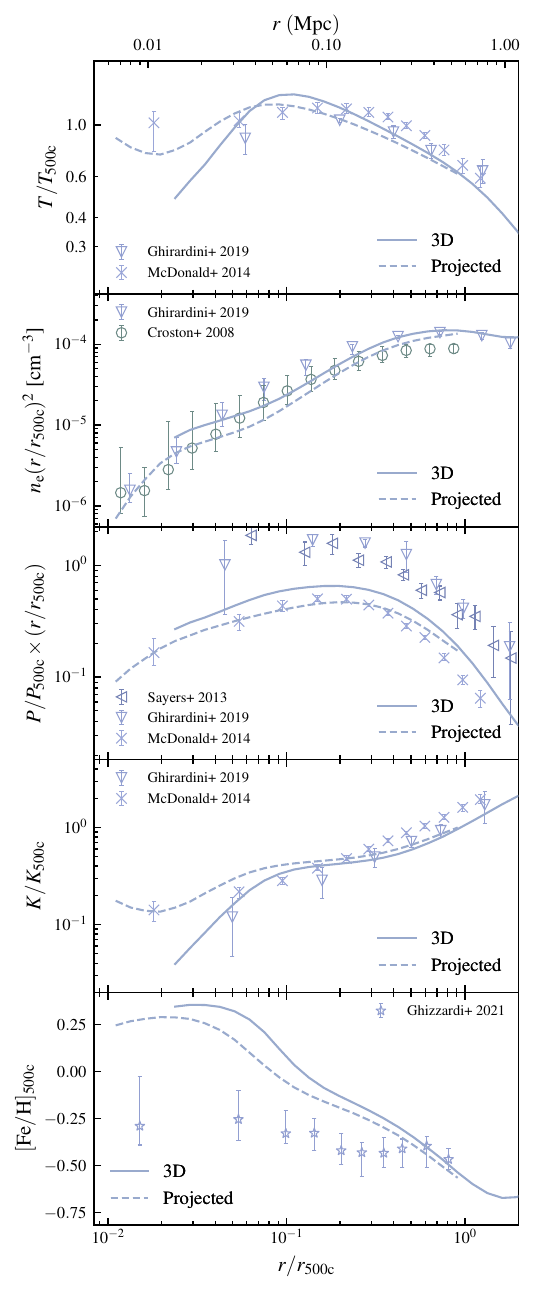}
    \caption{As Fig.~\ref{fig:weighting}, for clusters with mass ($M_{\rm 500c} = 10^{14.5}-10^{15} \mathrm{M_\odot}$), but showing 3D (solid line) and projected (dashed line) profiles for X-ray weighting at $z = 0$, compared with de-projected observational results.}
    \label{fig:projection}
\end{figure}

In Figure \ref{fig:projection} we compare 3D and projected profiles. We see large differences, especially in the core. Projected profiles from our simulations often seem to fit the deprojected observed profiles better than our 3D profiles. It would be of interest to test the deprojection algorithms used by observational studies on mock images from simulations to study systematic effects such as the influence of non-sphericitiy. This would then hopefully lead to a better understanding of these systematic effects which are especially important in the core, where we currently often see the largest differences between observed and simulated profiles. At larger radii the differences shrink, but remain non-negligible.

\section{Resolution} \label{sec:appendix_convergence}
\begin{figure}
    \centering
    \includegraphics{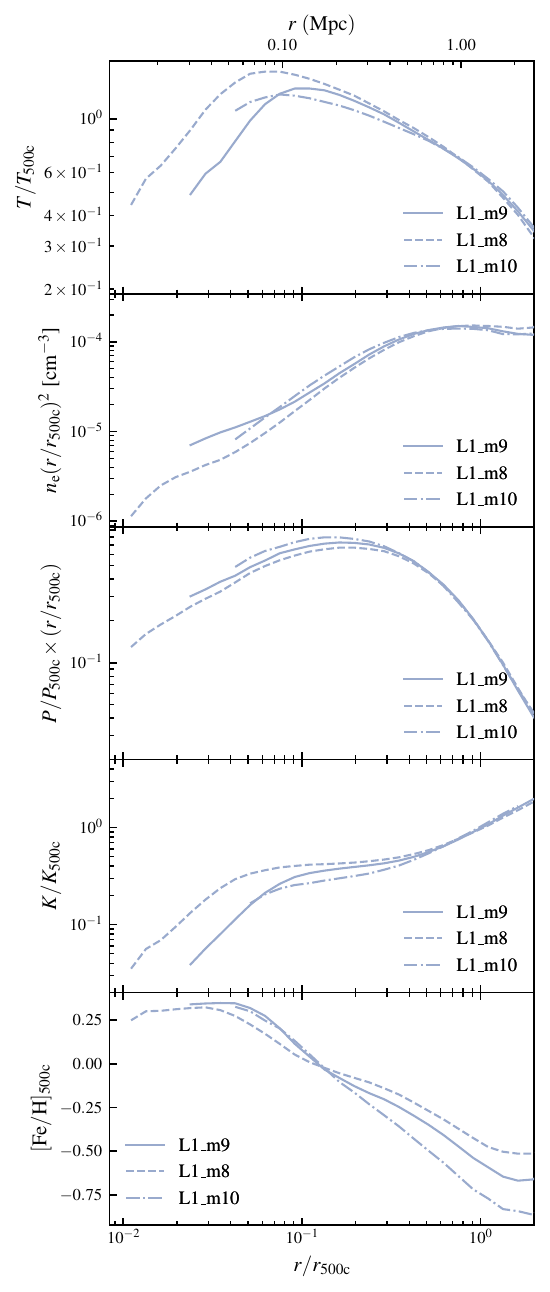}
    \caption{As Fig.~\ref{fig:weighting}, for clusters with mass ($M_{\rm 500c} = 10^{14.5}-10^{15} \mathrm{M_\odot}$), but showing different numerical resolutions. Differences are small at large radii, but tend to increase towards smaller radii.}
    \label{fig:resolution}
\end{figure}

We test the convergence properties of our simulations for our fiducial mass bin $M_{\rm 500c} = 10^{14.5} - 10^{15.0} ~ \mathrm{M_\odot}$, and show the results in Figure~\ref{fig:resolution}. We compare the three simulations L1\_m8, L1\_m9 and L1\_m10, which use the same box size but differ in mass (spatial) resolution by factors of 8 (2) (see Table~\ref{table:variations}). It should be kept in mind that although all resolutions have been calibrated to the observed low-redshift galaxy mass function and cluster gas fractions, they are not the same physical models.

For the pressure profile the resolution has a negligible effect. The temperature profile shows converged results at large radii, while at small radii the peak of the temperature shifts inwards and increases slightly in normalisation for the highest resolution. Something similar can be seen for the entropy profile, where the plateau extends to smaller radii for L1\_m8. The density profile shows a slight suppression at $r < R_{\rm 500c}$ but only for the highest resolution. The metallicity profiles are somewhat shallower for increasing resolution. In all cases, at small radii the highest resolution model is closer to the observations than an extrapolation of the intermediate resolution model.

\section{Normalisation}

\subsection{Pressure} \label{pressure_derivation}
We can derive the expected pressure at $r_{\rm 500,c}$ by starting from the definition for physical electron pressure and writing down some other standard virial relations
\begin{align}
    P_{\rm 500c} &= n_{\rm e, 500c} k_{\rm B} T_{\rm 500c} \label{eq:pressure_def} \\
    k_{\rm B} T_{\rm 500c} &= \frac{\mu m_{\rm p} G M_{\rm 500c}}{2 R_{\rm 500c}} \label{eq:kbT500}\\
    n_{\rm e, 500c} &= \frac{\rho_{\rm g, 500c}}{\mu_{\rm e} m_{\rm p}} \label{eq:ne500}\\
    \rho_{\rm g, 500c} &= 500 f_{\rm B} \rho_{\rm c}(z) \\
    \rho_{\rm c} &= \frac{3H^2(z)}{8\pi G} \label{eq:critical_density}
\end{align}
with $k_{\rm B}$ the Boltzmann constant, $m_{\rm p}$ the proton mass, $G$ the gravitational constant, $\mu_{\rm e}$ the mean particle mass per free electron, $f_{\rm B}$ the cosmic baryon mass fraction, and $H(z)$ the Hubble parameter.

\noindent Combining these relations, we can write eq.~\ref{eq:pressure_def} as

\begin{equation}\label{eq:press_inter_1}  
    n_{\rm e, 500c} k_{\rm B} T_{\rm 500c} = \frac{1500}{16\pi} f_{\rm B} H^2(z) \frac{\mu}{\mu_{\rm e}} \frac{M_{\rm 500c}}{R_{\rm 500c}} 
\end{equation}
We now need an expression for $R_{\rm 500c}$
\begin{align}
    R_{\rm 500c} &= \left( \frac{\frac{3}{4\pi} M_{\rm 500c}}{500 \rho_{\rm c}(z)} \right)^{1/3} \\
    &= \left( \frac{2G}{500} \frac{M_{\rm 500c}}{H^2(z)} \right)^{1/3} \label{eq:r500c} \, ,
\end{align}
where we used eq.~\ref{eq:critical_density}.
Putting this into eq.~\ref{eq:press_inter_1}, we obtain
\begin{equation}
    n_{\rm e, 500c} k_{\rm B} T_{\rm 500c} = \frac{1500}{16\pi} f_{\rm B} H^2(z) \frac{\mu}{\mu_{\rm e}} M_{\rm 500c}^{2/3} \left( \frac{500 H^2(z)}{2G} \right)^{1/3} \, .
\end{equation}
Taking part of the right-hand-side and rewriting gives
\begin{equation}
    \frac{1500 H^2(z)}{16 \pi} \left( \frac{500 H^2(z)}{2G} \right)^{1/3} = \frac{3}{8\pi} \left( \frac{500}{2G^{1/4}} H^2(z) \right)^{4/3} \, .
\end{equation}
Putting it all together gives an expression for the pressure at $R_{\rm 500c}$
\begin{equation}
    P_{500c} = \frac{3}{8\pi} \left( \frac{500}{2G^{1/4}} H^2(z) \right)^{4/3} f_{\rm B} \frac{\mu}{\mu_{\rm e}} M_{\rm 500c}^{2/3} \, .
\end{equation}

\subsection{Temperature} \label{temperature_derivation}
Similarly to the pressure, we can derive the temperature for an isothermal sphere at $R_{\rm 500c}$ by combining equations \ref{eq:kbT500} and \ref{eq:r500c}
\begin{equation}
    T_{\rm 500c} = \frac{\mu m_{\rm p}}{2k_{\rm B}} \left( \frac{500 G^2}{2} \right)^{1/3} M_{\rm 500c}^{2/3} H^{2/3}(z). \label{eq:T500c}
\end{equation}

\subsection{Entropy} \label{entropy_derivation}
For the entropy, we start from the definition of the physical entropy 
\begin{equation}
    K_{\rm 500c} = \frac{k_{\rm B}T_{\rm 500c}}{n_{\rm e, 500c}^{2/3}},
\end{equation}
and fill in the expressions for $T_{500}$ (eq.~\ref{eq:T500c}) and $n_{\rm e, 500c}$ (eq.~\ref{eq:ne500})
\begin{align}
    K_{\rm 500c} 
    &= \frac{\frac{\mu m_{\rm p}}{2} \left( \frac{500 G^2}{2} \right)^{1/3} M_{\rm 500c}^{2/3} H^{2/3}(z)}{\left (\frac{1500 f_{\rm B} H^2(z)}{8\pi G \mu_{\rm e} m_{\rm p}}\right )^{2/3}}\\
    &= M_{\rm 500c}^{2/3} H^{-2/3}(z) \frac{\left( \mu^3 \mu_{\rm e}^2 m_{\rm p}^5 G^4 \right)^{1/3}}{5 (3/\pi)^{2/3}f_{\rm B}^{2/3}}
\end{align} 
%%%%%%%%%%%%%%%%%%%%%%%%%%%%%%%%%%%%%%%%%%%%%%%%%%

\section{Different cool-core criteria} \label{sec:appendix_CC_criteria}
\begin{figure}
	\includegraphics[width=\linewidth]{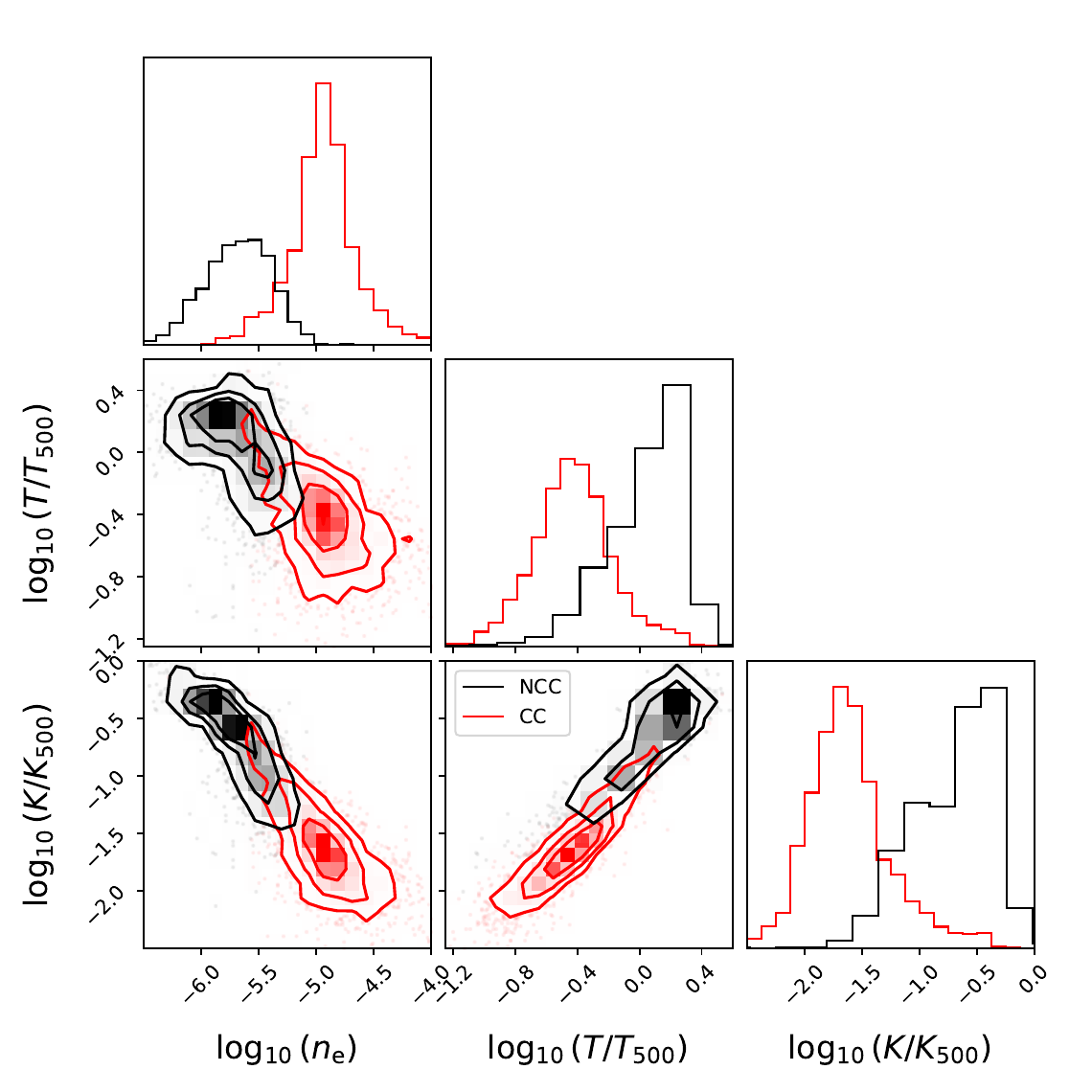}
    \caption{Separation between CC (red) and NCC (black) clusters, identified by their central cooling times, shown for different CC metrics, the central density, normalised temperature and normalised entropy.}
    \label{fig:CC_NCC_metrics}
\end{figure}

In the main text we have described a cool-core selection based on the central cooling time $t_{\rm cool}(r < 0.048~R_{\rm 500c})$. However, this quantity is not always available in observations as it requires detailed knowledge of the thermodynamic state of the cluster core. Instead, entropy or density thresholds are sometimes used. Apart from that, CC were originally identified as a population of objects with distinctly colder cores. Fig. \ref{fig:CC_NCC_metrics} compares the central ($r < 0.048~R_{\rm 500c}$) density, temperature, and entropy for the CC and NCC populations that were classified by their central cooling time.  For all the thermodynamic cool-core metrics, the two populations clearly separate, with CC objects having higher central density, and lower central entropy and temperature.

% Don't change these lines
\bsp	% typesetting comment
\label{lastpage}
\end{document}